\documentclass[12pt]{article}
\usepackage{amscd}
\usepackage{amssymb}
\usepackage{amsthm}
\usepackage{amsmath}
\usepackage{amsxtra}
\DeclareMathOperator{\Hom}{Hom}

\begin{document}
\title{Multiplicity, Invariants and Tensor Product Decompositions of
Tame Representations of U$\mathbf{ (\infty)}$}
\author{R. Michael Howe\\ 
        \small  Department of Mathematics\\
        \small  University of Wisconsin-Eau Claire\\
        \small  Eau Claire, Wisconsin \ 54702 \\
	\small  email: hower@uwec.edu \\
		\and
Tuong Ton-That\\
        \small  Department of Mathematics\\
        \small  University of Iowa\\
        \small  Iowa City, Iowa \ 52242 \\
	 \small  email: tonthat@math.uiowa.edu}

\bibliographystyle{plain}

\newtheorem{prop}{Proposition}[section]
\newtheorem{df}[prop]{Definition}
\newtheorem{thm}[prop]{Theorem}
\newtheorem{lemma}[prop]{Lemma}
\newtheorem{rmk}[prop]{Remark}
\newtheorem{cor}[prop]{Corollary}
\renewcommand{\thesection}{\Roman{section}}
\newlength{\li} \setlength{\li}{12pt}
\newcommand{\singlespace}{\baselineskip 1.0\li}
\newcommand{\doublespace}{\baselineskip 2.0\li}
\newcommand{\F}{ \mathcal{ F }}
\newcommand{\D}{ \mathcal{ D }}
\newcommand{\I}{ \mathcal{ I }}
\newcommand{\PP}{ {\mathcal P }}

\newcommand{\dlim}{ \underrightarrow{\lim}}
\newcommand{\ilim}{ \underleftarrow{\lim}}
\newcommand{\Finfty}{ \mathcal{ F_{\infty}}}
\newcommand{\C}{ \mathbb{ C }}
\newcommand{\N}{ \mathbb{ N }}
\newcommand{\R}{ \mathbb{ R }}

\newcommand{\Iinfn}{ \mathbb{ I^{\infty}_{N}}}
\newcommand{\Zplus}{ \mathbf{Z^{+} }}
\newcommand{\Zminus}{ \mathbf{Z^{-} }}
\newcommand{\DD}{ \mathbf{ D }}

\newcommand{\zeroinfty}{ \overrightarrow{0}}
\newcommand{\zeroneginfty}{ \overleftarrow{0}}

\newcommand{\rhocheck}{\rho\spcheck}
\newcommand{\Rcheck}{R\spcheck}
\newcommand{\gcheck}{g\spcheck}
\newcommand{\fmax}{f_{max}}
\newcommand{\fmin}{f_{min}}

\doublespace

\maketitle

\begin{abstract}

The structure of $r$-fold tensor products of irreducible tame
representations of $U(\infty) = \varinjlim U(n)$ are described,
versions of contragredient representations and invariants are realized,
and methods of calculating
multiplicities, Clebsch-Gordan and Racah coefficients are given using
 invariant theory on Bargmann-Segal-Fock spaces.

\end{abstract}

 
\section{Introduction}

        Let $G_{k}$ and $G_{k}^{\C}$ denote the unitary group and the
general linear group, respectively.  Then the {\em inductive limits}
        \[ G_{\infty}= \varinjlim \ G_{k} = \bigcup_{k=1}^{\infty} G_{k} \]
and
  \[ G_{\infty}^{\C}= \varinjlim \  G_{k}^{\C} = \bigcup_{k=1}^{\infty} G_{k}^{\C} \]
may be defined as follows:
\[ G_{\infty}^{\C}= \{ g=(g_{ij})_{i,j \in \N} \ | \ g \ \text{is invertible
and all but a finite number of} \ g_{ij}-\delta_{ij}=0 \}  \]
and
\[ G_{\infty}= \{u \in  G_{\infty}^{\C} \ | \ u^{*}=u^{-1} \}. \]
Representation theory of $G_{\infty}$ and  $G_{\infty}^{\C}$ was first
studied by I. Segal in \cite{se}, then by A. Kirillov in \cite{ki}, followed
by S. Stratila and D. Voiculescu in \cite{stvo}, D. Pickerell in \cite{pi},
G. Ol'shanskii in \cite{ol3} \cite{ol1} \cite{ol2}, I. Gelfand and M.
Graev in
\cite{gegr}, and
V. Kac in \cite{ka}.  This list is certainly not exhaustive, and the most
complete list of references can be found in the comprehensive and important
work of Ol'shanskii.

        Following Ol'shanskii we call a unitary representation of
$G_{\infty}$ {\em tame} if it is continuous in the group topology in which
the descending chain of subgroups of the type
        $\{ \left(
                \begin{smallmatrix}
                        1_{k} & 0 \\
                        0  & *
                \end{smallmatrix}  \right)  \}, $
$k = 1,2,3,\ldots$ constitutes a fundamental system of neighborhoods of the
identity $1_{\infty}$.  Assume that for each $k$ a unitary representation
$(R_{k}, H_{k})$ of $G_{k}$ is given and an isometric embedding (of Hilbert
spaces) $i_{k+1}^{k}: H_{k} \longrightarrow H_{k+1}$ commuting with the
action of $G_{k}$
(i.e., $i_{k+1}^{k} \circ R_{k} (u) = R_{k+1}(u) \circ i_{k+1}^{k})$
is given.  If $H_{\infty}$ denotes the Hilbert space completion of
$\bigcup_{k=1}^{\infty}\ H_{k}$, then there exists uniquely a unitary
representation $R_{\infty}$ of $G_{\infty}$ on $H_{\infty}$ defined by
        \[ R_{\infty} (u) f = R_{k}(u)f \ \ \text{if \ $u \in G_{k}$
                and $f \in H_{k}$ } . \]
The representation $(R_{\infty},H_{\infty})$ is called the
{\em inductive limit} of the sequence $(R_{k}, H_{k})$, and we have the
following Theorem (see \cite{ol3} for a proof).
\begin{thm}  \label{olthm}
        If the representations $(R_{k},H_{k})$ are all irreducible then the
inductive limit $(R_{\infty}, H_{\infty})$ is also irreducible.
\end{thm}

        Let
        \[ \lambda_{G_{k}}=(m_{1}, \ldots, m_{k}),\ \
                m_{1} \geq m_{2} \geq \ldots \geq m_{k} \geq 0, \ \
                m_{i} \in \N \cup \, \{ 0 \}, \ \ i=1 \ldots k.  \]
then  Ol'shanskii  proved the following
\begin{thm}
        All unitary irreducible tame representations of $G_{\infty}$ are
the inductive limits of the sequences of the form
$ \{ \rho_{\lambda}, V^{\lambda_{G_{k}}} \} $, where in each
$(\lambda)=(m_{1}, m_{2}, \ldots )$ the $m_{i}$ are equal to $0$ for
sufficiently large $i$.
\end{thm}

        It follows from `Weyl's unitarian trick' that all irreducible
tame representations of $G_{\infty}$ are inductive limits of sequences of
the form  $ \{ \rho_{\lambda}, V^{\lambda_{G_{k}}} \} $. Following Ol'shanskii
a representation of $G_{\infty}$ is called {\em holomorphic} if it is a
direct sum (of any number) of irreducible tame representations.

        In this paper we consider the problem of decomposing an $r$-fold
tensor product of unitary irreducible tame representations of $G_{\infty}$.
Such a problem was investigated in \cite{ki} and \cite{gegr} for the
simplest type of tame irreducible representations, namely the fundamental
(or principal) ones.  In light of the recent interest in Physics in the
representation theory of $U_{\infty}$ it is natural to consider such an
important problem in this theory.

        The general problem can be stated as follows:  Given $r$ tame
irreducible $(G_{\infty}, V^{(\lambda_{i})^{\infty}} )$ modules, choose a
basis $ | \lambda_{i}, \xi_{i} \rangle$ for each $i$  (such a basis always
exists, for example, the generalized Gelfand-\v{Z}etlin basis given in
\cite{gegr}, but we do not limit ourselves only to this basis).  Form the
$r$-fold tensor product
$(\lambda_{1})^{\infty} \otimes \cdots \otimes (\lambda_{r})^{\infty}$
and calculate the number of times the irreducible representation
$(\lambda)^{\infty}$ occurs in the tensor product.  The first method to
compute this multiplicity is to observe that the spectral decomposition
(or Clebsch-Gordan series) stabilizes for $k$ sufficiently large and then
apply the Weyl determinant formula for $U(k)$ for sufficiently large $k$.
This fact is proved rigorously as a theorem in Section \ref{stability}.  In
\cite{kt} it was shown that this multiplicity (for $SU(k)$) can be computed
as solutions of Diophantine equations arising from the invariants of $SU(k)$.
The first part of our program which is similar to the strategy given in
\cite{kt} is as follows:  instead of computing the multiplicity of
$(\lambda)^{\infty}$ in the tensor product
$(\lambda_{1})^{\infty} \otimes \cdots (\lambda_{r})^{\infty}$ we look at
what is equivalent, the multiplicity of the identity representation in the
augmented tensor product
$(\lambda_{1})^{\infty} \otimes \cdots \otimes (\lambda_{r})^{\infty}
\otimes (\lambda \spcheck)^{\infty}$ where $(\lambda \spcheck)^{\infty}$ is
the contragredient representation of $(\lambda)^{\infty}$.  But with this
approach we are facing two major difficulties.  The first one pertains to
the contragredient representation $(\lambda \spcheck)^{\infty}$: as it is well known
(see e.g. \cite{ol2}) an irreducible
$(G_{\infty}, V^{(\lambda)^{\infty}} )$-module can be realized as a subspace
of a generalized Bargmann-Segal-Fock space in $n \times \infty$ complex
variables (see Section \ref{prelim} for this realization), but it is not
known whether the irreducible
$(G_{\infty}, V^{(\lambda \spcheck)^{\infty}} )$-module is realizable
likewise.  We prove in Section \ref{contra} that by `twisting' the action of
the contragredient representation and by using an appropriate embedding of
 Bargmann-Segal-Fock spaces
$\F_{n \times k} \subset \F_{n \times (k+1)} \subset \cdots $   the
$(G_{\infty}, V^{(\lambda \spcheck)^{\infty}} )$-module can also be realized
as a submodule of a  Bargmann-Segal-Fock space $\F_{n \times \infty}$.  The
notable difference is that the signature of $(\lambda \spcheck)^{\infty}$ is
characterized by the {\em lowest weight} instead of the highest weight and we
will be dealing with {\em lowest weight vectors} instead of highest weight
vectors as in the case $(G_{\infty}, V^{(\lambda)^{\infty}} )$.  Another
difficulty is that, realized as a submodule of the
Bargmann-Segal-Fock space $\F_{n \times \infty}$, it is not clear that the
tensor product $(\lambda_{1})^{\infty} \otimes \cdots \otimes
(\lambda_{r})^{\infty} \otimes (\lambda \spcheck)^{\infty}$ considered as a
$G_{\infty}$-module is a holomorphic representation; in particular, the
identity representation might not occur in this tensor product.  Using a
general reciprocity theorem for holomorphic representations of some
infinite-dimensional groups (see \cite{ttone}) we show that the tensor
product $(\lambda_{1})^{\infty} \otimes \cdots \otimes
(\lambda_{r})^{\infty} \otimes (\lambda \spcheck)^{\infty}$ is indeed a
holomorphic representation and that the multiplicity of the identity
representation of $G_{\infty}$ in this augmented tensor product is indeed
equal to the multiplicity of $(\lambda)^{\infty}$ in the tensor product
$(\lambda_{1})^{\infty} \otimes \cdots (\lambda_{r})^{\infty}$.  Having
overcome this difficulty first, we still have to deal with a second major
difficulty; the generators
of $SU(k)$ used in \cite{kt} which are determinants of matrices of order
$k$ become unmanageable when $k$ is large; furthermore, at the limit as
$k \longrightarrow \infty$ these determinants are certainly not members of
$\F_{n \times \infty}$.  Both of these problems can be dealt with as
follows: instead of using the determinant-invariants of $SU(k)$ we use the
classical invariants of $U(k)$ which are generated by a system of
algebraically independent polynomials, but more importantly, the number
of these polynomials depends only on the tensor product
$(\lambda_{1})_{k} \otimes \cdots \otimes (\lambda_{r})_{k}
\otimes (\lambda \spcheck)_{k}$ and not on $k$; in fact, the problems
considered in \cite{kt} can be entirely solved using this new approach.
Next, it can be shown (see \cite{tttwo}) that when $k \longrightarrow \infty$
these invariants tend to their {\em inverse} or {\em projective limits}
which are infinite formal series of complex variables, but nevertheless
remain algebraically independent and generate all $G_{\infty}^{\C}$
(or $G_{\infty}$) invariants.  By analogy with the definition of {\em rigged
Hilbert Spaces} (c.f. e.g. "Generalized functions" by I. M. Gelfand and
G. E. Shilov, Vol 4, P. 106) these infinite formal series may be thought of
as differential operators.  Thus if $f \in \F_{n \times \infty}$ and $p$ is a
$G_{\infty}$-invariant then the inner product
        \[ \langle p , f \rangle = p(D)\overline{f(\bar{Z})}|_{Z=0} \]
makes perfect sense since $f \in \F_{n \times k}$ for some $k$ and those
terms in $p(D)$
 whose column indices are larger than $k$ simply evaluate to zero. With this new
interpretation of the $G_{\infty}$-invariants the method of computing
Clebsch-Gordan and Racah coefficients in \cite{kt} can be adapted to the
case of tensor products of $G_{\infty}$;  Actually, both the Diophantine
equations and the computations of Clebsch-Gordan coefficients are much
simpler, since the $G_{\infty}$-invariants are much simpler.

\section{Preliminaries} \label{prelim}

        Let $\C^{n \times k}$ denote the vector space of $n$ row by $k$ column
matrices over $\C$, the field of complex numbers. If $Z=(Z_{ij})$ is an element of
$\C^{n \times k}$, we let ${\bar Z}$ denote its complex conjugate, and write
        \[Z=X_{ij} + \sqrt{-1}\  Y_{ij}\,; \ \ 1 \leq i \leq n, \ 1
                \leq j \leq k .\]
If $dX_{ij}$
(resp. $dY_{ij}$) denotes the Lebesgue measure on $\R$ we let
        $dZ=\prod_{1\leq i \leq n,\  1 \leq j \leq k}\  dX_{ij}\, dY_{ij} $
denote the Lebesgue product measure on $\R^{nk}$.
Define a Gaussian measure $d \mu$ on $\C^{n \times k}$ by
        \[ d \mu (Z) = \pi^{-nk}\exp [ tr (Z {\bar Z}^t)] dZ \]
where $tr$ denotes the trace of a matrix.
A map $f:\C^{n \times k} \rightarrow \C$ is said to be
\emph{holomorphic square integrable} if it is holomorphic on the entire
domain $\C^{n \times k}$ and if
        \[\int_{\C^{n \times k}} |f(Z)|^{2} d \mu (Z) < \infty . \]
The holomorphic square integrable functions form a Hilbert space with
respect to the inner product
        \begin{equation}
           \label{innerprod}
                (f_{1},f_{2}) = \int_{\C^{n \times k}}
                f_{1} (Z) \overline{f_{2} (Z) } d \mu (Z) ,
        \end{equation}
of which the polynomial functions form a dense subspace.
The inner product (\ref{innerprod}) is equivalent to the inner product
\begin{equation}
           \label{innerprodb}
                \langle f_{1},f_{2} \rangle
                        =f_{1}(D) \overline{f_{2}(\bar{Z})}|_{Z=0}
        \end{equation}
where $f(D)$ is the
differential operator obtained by formally replacing
$Z_{ij}$ by the partial derivative $\partial / \partial Z_{ij}$.
We denote this Hilbert space by $\F_{k}=\F(\C^{n \times k})$.
The natural embedding of $\C^{n \times k}$ into $\C^{n \times (k + 1)}$
given by

\[   Z \mapsto \left( \begin{array}{cccc}
                                \ & \  & \ & 0 \\
                                 \ & Z & \ & \vdots \\
                                \ & \ & \ & 0
                                \end{array} \right)
                                \in \C^{n \times (k + 1)}  \]
induces an isometric embedding
        \[ i_{k+1}^{k}: \F_{k}  \longrightarrow \F_{k+1} \]
 so that the collection $ \{ \F_{k}, i_{k+1}^{k} \} $  forms a
directed system.
We
can then take the \emph{inductive limit} $\Finfty =
   \overline{ \varinjlim \F_{k} } $
(where the bar indicates closure with respect to the norm), with the natural
inclusion
        \[ i_{k}: \F_{k}  \longrightarrow \F_{\infty}. \]
Formally, elements of $\Finfty$ are realized as equivalence classes
$[f_{\alpha}]$, where \[f_{\alpha} \sim f_{\beta} \ \ \mathrm{whenever }
f_{\beta} = i_{\beta}^{\alpha} (f_{\alpha})\ \  \mathrm{and} \ \
 \alpha \leq \beta,  \ \ \   \alpha, \beta \in \N.\]
Since in our case we have  $\F_{k}  \subset \F_{k+1},$
we can realize this space as
         \[ \Finfty = \overline{\bigcup_{k=1}^{\infty}  \F_{k}} .\]
If $G_{k} = U(k)$ (or  $ U(k)^{\C} = GL(k, \C)$) we also have the
natural
inclusion
 \[ j_{k+1}^{k}:G_{k}\longrightarrow G_{k+1}\]
 given by
\begin{equation} \label{G}
                   g \mapsto \left( \begin{array}{cccc}
                                \ & \  & \ & 0 \\
                                 \ & g & \ & \vdots \\
                                \ & \ & \ & 0 \\
                                0   & \cdots& 0 & 1
                                \end{array} \right) \in G_{k+1}
\end{equation}
We can then take the inductive limit
 $G_{\infty} = \varinjlim G_{k}$, with the natural inclusion
$j_{k}:G_{k}\longrightarrow G_{\infty}$.
Again elements of $G_{\infty}$ are
formally defined as equivalence classes $[g_{k}]$, where we identify some
$g_{k} \in G_{k}$ with its inclusions into $G_{k+1}, \ G_{k+2}$, etc.
If we let $R_{k}$ denote the
representation of $G_{k}$ on $\F_{k}$ given by right translation
        \[ R_{k}(g)f(Z) = f(Zg), \ \ Z \in \C^{n \times k},\  g \in G_{k} .\]
Then the following diagram commutes
\begin{equation}  \label{cda}
\begin{CD}
        G_{k} \times \F_{k}       @>R_{k}>>         \F_{k} \\
        @Vj_{k+1}^{k}\times i_{k+1}^{k}VV       @VVi_{k+1}^{k}V \\
        G_{k+1} \times \F_{k+1}   @>>R_{k+1}>        \F_{k+1}
\end{CD}
\end{equation}
and so the representation $R=\varinjlim R_{k}$ of $G_{\infty}$ on $\F_{\infty}$
is well defined by
\begin{equation}  \label{welldef}
        R ( [g_{k}])[f_{k}]=[R_{k}(g_{k}) f_{k}],
\end{equation}
 called the
{\em inductive limit} of the  representations $R_{k}$
on $\Finfty$,
and we have commutativity of the diagram
\begin{equation*}  % \label{cd}
\begin{CD}
        G_{k} \times \F_{k}       @>R_{k}>>         \F_{k} \\
        @Vj_{k}\times i_{k}VV                       @VVi_{k}V \\
        G_{\infty} \times \F_{\infty}   @>>R>        \F_{\infty}
\end{CD}
\end{equation*}
%and $R$ is said to be the
%{\em inductive limit} of the  representations $R_{k}$ on $\Finfty$.

Let $\DD_{k} \subset G_{k}$ be the diagonal subgroup.
Let $\Zplus_{k} \subset G_{k}$ be the unipotent subgroup of upper triangular matrices
with
ones along the main diagonal and let $\Zminus_{k}$ be the analogous lower
triangular subgroup.
If $(M)= (M_{1}, \ldots, M_{k})$ is any collection of  integers, we define a
holomorphic character
 \begin{equation*}  %\label{pid}
         \pi^{(M)} (d) = d_{11}^{M_{1}} d_{22}^{M_{2}}
                 \ldots d_{kk}^{M_{k}} \ \ \ d \in \DD_{k}.
  \end{equation*}
In this context an element
$f \in \F_{k}$ is said to be a {\em weight vector} of the
representation $R_{k}$ with weight $(M)$ if
        \[ [R_{k}(d)f](Z)=f(Zd) = \pi^{(M)} (d) f(Z), \ \ \ \
                \forall \  d \in \DD_{k} .\]
If $f$ is a weight vector, and if
   \[ [R_{k}(\zeta)f](Z)=f(Z \zeta) =   f(Z), \ \ \ \   \forall  \
                 \zeta \in \Zplus_{k}\]
then $f \in \F_{k}$ is a  said to be a {\em highest weight vector} of the
representation $R_{k}$.  Similarly if  $f$ is a weight vector, and if
 \[ [R_{k}(\zeta)f](Z)=f(Z \zeta) =   f(Z),  \ \ \ \  \forall \ \
                 \zeta \in \Zminus_{k}\]
then $f \in \F_{k}$ is a  said to be a {\em lowest weight vector}  of the
representation $R_{k}$.
Since $G_{k}=U(k)$ or $GL(k,\C)$, each irreducible representation of $G_{k}$
in  $\F_{k}$ is finite dimensional and so admits a ``unique''
(up to multiplication by a nonzero scalar) highest weight vector with
highest weight
$(m)=(m_{1}, m_{2}, \ldots , m_{k})$, and a unique lowest weight vector with
lowest weight $(m_{k}, m_{k-1}, \ldots , m_{1})$.
This highest (or lowest) weight then characterizes each irreducible
representation of $G_{k}$, and is called the {\em signature} of the
representation. By the Borel-Weil theorem a necessary and sufficient
condition for $(m)$ to be the highest weight of an irreducible polynomial
 representation of $G_{k}$
on $\F_{k}$ is that
$m_{1}\geq m_{2} \geq \ldots \geq m_{k} \geq 0 $.  $V^{(m)}$ is then
cyclically generated as a
$G_{k}$-module by the action of $G_{k}$ on any one of its elements, in
particular
on its highest (or lowest) weight vector.
Let $B_{k} \subset GL(k,\C)$ be the Borel subgroup of lower triangular
 matrices
and for a $k$-tuple of non-negative integers
 $(m)=(m_{1}, m_{2}, \ldots , m_{k})$ we define a holomorphic character
  \begin{equation}  \label{pi}
         \pi^{(m)} (b) = b_{11}^{m_{1}} b_{22}^{m_{2}}
                 \ldots b_{kk}^{m_{k}} \ \ \ b \in B_{k}.
  \end{equation}
As a  consequence of the Borel-Weil theorem (see for example \cite{dz}), any
irreducible
holomorphic
representation of $G_{k}$ with
signature $(m)=(m_{1}, m_{2}, \ldots , m_{n})$ can be explicitly
realized as the representation $R_{k}$
on the subspace of polynomial
functions in $\F_{k} = \F (\C^{n \times k})$
which satisfy the covariant condition
\begin{equation} \label{covariant}
         f(bZ) = \pi^{(m)} (b) f(Z), \ \  \ \ \ b \in B_{n}.
\end{equation}
We denote this subspace by $V^{(m)}_{k}$, the restriction of $R_{k}$
to this subspace by $R_{k}^{(m)}$, and where necessary we explicitly
designate
this irreducible representation by the pair $(R_{k}^{(m)},V^{(m)}_{k})$.

        For each $k = 1,2,\ldots$, let $V_{k}$ be a subspace of  $\F_{k}$,
on which the
representation $R_{k}$
of $G_{k}$  is irreducible.  Suppose also that the following
 diagram commutes
\begin{equation}  \label{cdb}
\begin{CD}
        G_{k} \times V_{k}       @>R_{k}>>         V_{k} \\
        @Vj_{k+1}^{k}\times i_{k+1}^{k}VV       @VVi_{k+1}^{k}V \\
        G_{k+1} \times V_{k+1}   @>>R_{k+1}>        V_{k+1}
\end{CD}
\end{equation}
or equivalently, that the restriction of $R_{k+1}$ to $G_{k}$ contains a
representation equivalent to $R_{k}$.  In this case we write
$V_{k} \preceq V_{k+1}$, and it is well documented in the literature
that the representation $R = \varinjlim R_{k}$ of
$G_{\infty}$ on $V=\varinjlim \ V_{k}$ is also irreducible. (For detailed
expositions of inductive limit representations see \cite{gegr} \cite {ki}
\cite{ol3}.)

  If $V_{k}$ is an irreducible representation of $G_{k}$
with  signature
\[ (m)= (m_{1},m_{2},\ldots , m_{k}) \ \
\text{with  $\ m_{1} \geq m_{2} \geq \ldots \geq m_{k} \geq 0$,}  \]
and if  $V_{k+1}$ is
an irreducible representation of $G_{k+1}$ with signature
$(h)=(h_{1},h_{2},\ldots , h_{n},h_{n+1})$
it is also well known
that
$V_{k} \preceq V_{k+1}$ or equivalently written $(m) \preceq (h)$
 if and
only if
   \[ h_{i}\geq m_{i} \geq h_{i+1},  \ \ i=1,\ldots, k. \]
In particular, if $(m_{1},m_{2},\ldots , m_{k})$ is the signature of an
irreducible representation of $G_{k}$ and $(m_{1},m_{2},\ldots , m_{k}, 0
)$
is the signature of an irreducible representation of $G_{k+1}$, then
\[(m_{1},m_{2},\ldots , m_{k})
                \preceq (m_{1},m_{2},\ldots , m_{k}, 0) \]
and it is easy to see that if $\fmax$ is a highest weight vector for an
irreducible representation of $G_{k}$ with highest weight
$(m_{1},m_{2},\ldots , m_{k})$ then $\fmax$ is also a highest weight
vector of the irreducible representation of $G_{k+1}$ with highest weight
$(m_{1},m_{2},\ldots , m_{k}, 0)$.
We denote the inductive limit of the representations

        \[(m_{1},m_{2},\ldots , m_{k})
                \preceq (m_{1},m_{2},\ldots , m_{k}, 0)
                        \preceq (m_{1},m_{2},\ldots , m_{k}, 0,0)
                                \preceq \cdots \]
 by
        \[ (m_{1},m_{2},\ldots , m_{k}, 0 ,\ldots)=
                (m_{1},m_{2},\ldots , m_{k}, \zeroinfty) =
                        (m)^{\infty} \]
and realize this representation as the submodule of $\F_{\infty}$ generated
by the action of
$G_{\infty}$ on this highest weight vector.
In the sequel we may also require more explicit notation:
If $ \underbrace{(m_{1},m_{2},\ldots , m_{l},0,  \ldots, 0)}_{k}$
is the signature of an irreducible
representation of $G_{k}$ call the integers
$m_{1},m_{2},\ldots , m_{l},0$ the {\em entries}, we say $l$ is the
{\em length} of the signature
(i.e., if the signature has at most $l$ non-zero entries) and write
\[ \underbrace{(m_{1},m_{2},\ldots , m_{l},0, \ldots, 0)}_{k}= (m)^{k}_{l} \]
or just $(m)^{k}$ if it is unnecessary to specify the length.
With this notation we denote the signature of the
inductive limit of the representations
  \[(m_{1},m_{2},\ldots , m_{l})\preceq (m_{1},m_{2},\ldots , m_{l},0)
        \preceq
(m_{1},m_{2},\ldots , m_{l},0,0)\preceq \cdots \]
 \[ = (m)^{k}_{l} \preceq (m)^{k+1}_{l} \preceq (m)^{k+2}_{l}
                        \preceq \cdots \]

by
 \[(m)_{l}^{\infty} = (m_{1},m_{2},\ldots , m_{l}, 0,0,0,\ldots )
=(m_{1},m_{2},\ldots , m_{l}, \zeroinfty ).\]

\section{Stability of spectral decompositions} \label{stability}

We motivate this section with the following example.  It is readily computed
using one of the standard formulae (for example \cite{dz}) that the tensor
product of the irreducible representations of $G_{2} \ (= U(2)$ or $GL(2, \C))$
 decomposes as a
direct sum
\begin{multline} \label{u2rep}
        (1,0) \otimes (2,0) \otimes (2,0) \otimes(3,0) \\
        =(8,0)+3(7,1)+5(6,2)+5(5,3)+2(4,4)
\end{multline}
and the tensor product of irreducible representations of $G_{4}$
\begin{multline}  \label{u4rep}
        (1,0,0,0) \otimes (2,0,0,0) \otimes (2,0,0,0) \otimes(3,0,0,0) \\
        =(8,0,0,0)+3(7,1,0,0)+5(6,2,0,0)+5(5,3,0,0)+2(4,4,0,0)\\
        + 3(6,1,1,0) + 6(5,2,1,0) + 5(4,3,1,0)+3(4,2,2,0) + 2(3,3,2,0)\\
        +(5,1,1,1)+2(4,2,1,1)+(3,3,1,1)+(3,2,2,1)
\end{multline}
But notice that the first line of (\ref{u4rep}) is just (\ref{u2rep}),
the spectrum of $G_{2}$ embedded in
the spectrum of $G_{4}$.  In this case we say that  the spectrum of
$G_{4}$ {\em contains } the spectrum of $G_{2}$, or that the spectrum of
$G_{2}$ {\em appears in} the spectrum of $G_{4}$.  Furthermore, it is
 routine to check
that the spectral decomposition of irreducible representations of $G_{5}$
is given by
\begin{equation} \label{gfive}
\begin{aligned}
 \  &(1,0,0,0,0) \otimes (2,0,0,0,0) \otimes (2,0,0,0,0)
        \otimes(3,0,0,0,0) \  \\
        & =(8,0,0,0,0)+3(7,1,0,0,0)+5(6,2,0,0,0)+5(5,3,0,0,0)+2(4,4,0,0,0)\\
  & \ \  + 3(6,1,1,0,0) + 6(5,2,1,0,0) + 5(4,3,1,0,0)+3(4,2,2,0,0)\\
 & \ \ + 2(3,3,2,0,0) +(5,1,1,1,0)+2(4,2,1,1,0)+(3,3,1,1,0)\\
        & \ \ +(3,2,2,1,0)
\end{aligned}
\end{equation}
and  that the corresponding spectral decompositions of
$G_{6},\ G_{7},\ldots$ are the same, i.e. composed entirely of
the embedding of the
spectrum of
$G_{4}$.  In this case
we say the spectral decomposition {\em stabilizes}.

\begin{prop}
        If $(\alpha)^{k}_{k} = (\alpha_{1}, \alpha_{2}, \ldots ,\alpha_{k})$
 and
$(\beta)^{k}_{k} = (\beta_{1}, \ldots , \beta_{k})$
are the signatures of irreducible representations of $G_{k}$, and if
 $(\alpha)^{k+1}_{k} = (\alpha_{1}, \alpha_{2}, \ldots ,\alpha_{k},0)$
 and
$(\beta)^{k+1}_{k} = (\beta_{1}, \ldots , \beta_{k},0)$ are the
signatures of irreducible representations of $G_{k+1}$,
then the spectrum of
  $(\alpha)^{k}_{k} \otimes (\beta)^{k}_{k}$ appears in the spectrum of
  $ (\alpha)^{k+1}_{k} \otimes (\beta)^{k+1}_{k}$.  Furthermore, the
spectrum
of $(\alpha)^{K}_{k} \otimes (\beta)^{K}_{k}$ stabilizes for $K$ sufficiently
large.

\end{prop}
\begin{proof}
        We first note that, in the special case where
        $(\alpha)^{k}_{1} = \underbrace{(\alpha_{1}, 0 \ldots , 0)}_{k}$
 by \cite{dz}  the spectral decomposition of
        $(\alpha)^{k}_{1} \otimes (\beta)^{k}_{k}$
is given by the `Weyl formula', which is equivalent to  applying the
multiplier $\Gamma_{\alpha_{1}}$ to the signature
$(\beta)^{k}_{k}= (\beta_{1}, \ldots , \beta_{k})$
where
        \[ \Gamma_{\alpha_{1}} (\beta_{1}, \ldots , \beta_{k}) =
        \sum_{\substack{\nu_{1}+ \ldots + \nu_{k}=\alpha_{1}\\
                 0 \leq \nu_{i+1} \leq s_{i}}}
                (\beta_{1}+ \nu_{1}, \ldots , \beta_{k}+ \nu_{k})  \]
Here, and in what follows of this proof, the Weyl formula also
requires the condition that
$0 \leq \nu_{i+1} \leq s_{i}$ where $s_{i} = m_{i}-m_{i+1}$, and we will refer
to a multiplier of this type as a {\em simple multiplier}.

Now applying this simple multiplier to the signature $(\beta)^{k+1}_{k}$
we have
   \[ \Gamma_{\alpha_{1}} (\beta_{1}, \ldots , \beta_{k},0) =
        \sum_{\substack{\nu_{1}+ \ldots + \nu_{k}+ \nu_{k+1}=\alpha_{1} \\
                 0 \leq \nu_{i+1} \leq s_{i}}}
        (\beta_{1}+ \nu_{1}, \ldots , \beta_{k}+ \nu_{k},0+\nu_{k+1})  \]

\begin{multline*}
 = \sum_{\substack{\nu_{1}+ \ldots +  \nu_{k+1}=\alpha_{1} \\
                \nu_{k+1} = 0}}
                (\beta_{1}+ \nu_{1}, \ldots , \beta_{k}+ \nu_{k},
                        0+\nu_{k+1})
 \\ +\sum_{\substack{\nu_{1}+ \ldots +  \nu_{k+1}=\alpha_{1} \\
                \nu_{k+1} \neq 0}}
        (\beta_{1}+ \nu_{1}, \ldots , \beta_{k}+ \nu_{k},0+\nu_{k+1})
\end{multline*}
But the first sum, with $\nu_{k+1} = 0$, is just the spectrum of
        $(\alpha)^{k}_{1} \otimes (\beta)^{k}_{k}$ contained in the spectrum
of $(\alpha)^{k+1}_{1} \otimes (\beta)^{k+1}_{k}$

        We next note that a similar situation occurs when we apply a second
simple multiplier $\Gamma_{\alpha_{2}}$ to the above spectral decomposition
         $(\alpha)^{k+1}_{1} \otimes (\beta)^{k+1}_{k}$.
That is, the sums are grouped into those terms whose last entry is zero, and
those terms
whose last entry is non-zero;

 \[ \Gamma_{\alpha_{2}} \left[(\alpha)^{k+1}_{1} \otimes
                (\beta)^{k+1}_{k} \right]
        =  \Gamma_{\alpha_{2}}(\Gamma_{\alpha_{1}}
                (\beta_{1}, \ldots , \beta_{k},0))\]

 \begin{multline*}
=\Gamma_{\alpha_{2}} \left[
  \sum_{\substack{\nu_{1}+ \ldots + \nu_{k}+ \nu_{k+1}=\alpha_{1} \\
        \nu_{k+1} = 0}}
                (\beta_{1}+ \nu_{1}, \ldots , \beta_{k}+ \nu_{k},
                        0+\nu_{k+1}) \right. \\
 \left. +\sum_{\substack{\nu_{1}+ \ldots + \nu_{k}+ \nu_{k+1}=\alpha_{1}\\
                        \nu_{k+1} \neq 0}}
  (\beta_{1}+ \nu_{1}, \ldots , \beta_{k}+ \nu_{k},0+\nu_{k+1}) \right]
 \end{multline*}

\begin{multline*}
= \sum_{\substack{\nu_{1}+ \ldots + \nu_{k}+ \nu_{k+1}=\alpha_{1}\\
                \nu_{k+1} = 0}}
     \Gamma_{\alpha_{2}}(\beta_{1}+ \nu_{1}, \ldots , \beta_{k}+ \nu_{k},0 )  \\
  +\sum_{\substack{\nu_{1}+ \ldots + \nu_{k}+ \nu_{k+1}=\alpha_{1} \\
                \nu_{k+1} \neq 0}}
 \Gamma_{\alpha_{2}}
 (\beta_{1}+ \nu_{1}, \ldots , \beta_{k}+ \nu_{k},0+\nu_{k+1})
 \end{multline*}

\begin{multline*}
= \sum_{\substack{\nu_{1}+ \ldots +   \nu_{k+1}=\alpha_{1} \\
                \nu_{k+1} = 0}}
\     \sum_{\substack{\mu_{1}+ \ldots  + \mu_{n+1}=\alpha_{2}\\
                \mu_{k+1} = 0}}
(\beta_{1}+ \nu_{1}+ \mu_{1}, \ldots , \beta_{k}+ \nu_{k}+ \mu_{k},0 )  \\
                                                                       \\
+\sum \text{(other terms involving signatures whose last entry is non-zero)}
 \end{multline*}

        We then extend this idea to the general case where the spectral
decomposition of  $(\alpha)^{k}_{k} \otimes (\beta)^{k}_{k}$ is given by
applying the multiplier $\Gamma_{\alpha_{k}^{k}}$ to the signature
$\beta_{k}^{k}$ where $\Gamma_{\alpha_{k}^{k}}$ is a
{\em compound multiplier} computed as the
`Weyl Determinant' \cite{dz};

\begin{equation*}  %\label{weyldet}
        \Gamma_{\alpha_{k}^{k}} = \left|
        \begin{array}{llll}
                \Gamma_{\alpha_{1}} & \Gamma_{\alpha_{1}+1} & \cdots
                        & \Gamma_{\alpha_{1}+(k-1)}\\
                \Gamma_{\alpha_{2}-1} & \Gamma_{\alpha_{2}}& \cdots
                        & \Gamma_{\alpha_{2}+(k-2)}\\
                \vdots &  \cdots & \  & \vdots \\
        \Gamma_{\alpha_{k}-(k-1)} & \Gamma_{\alpha_{k}-(k-2)} & \cdots
                        & \Gamma_{\alpha_{k}}
        \end{array} \right|
\end{equation*}
        Here the simple multipliers $\Gamma_{\alpha}$ are regarded as
 permutable operators and the
determinant is expanded in the usual way, with
$\Gamma_{0}=1$ and $\Gamma_{\alpha}=0$ for $\alpha<0$.
From this last statement it is obvious that the compound multiplier
$\Gamma_{\alpha_{k}^{k+1}}$ is equal to $\Gamma_{\alpha_{k}^{k}}$ since the
$k+1^{st}$ row used to compute the determinant corresponding to
$\Gamma_{\alpha_{k}^{k+1}}$is just
$(0,\ldots,0,1)$.

        Now for notational convenience we set the simple multiplier
 $\Gamma_{i,j}=\Gamma_{\alpha_{i}-(i-j)}$ and using the usual formula for
determinant (summing over $S_{k}$, the symmetric group on $k$
symbols) we have

   \[\Gamma_{\alpha_{k}^{k+1}}
        =\Gamma_{\alpha_{k}^{k}}= \sum_{\sigma \in S_{k}} sgn(\sigma)
                \Gamma_{\alpha_{1} \sigma(1)} \cdots
                \Gamma_{\alpha_{k} \sigma(k)}   \]

So
\begin{multline}
   \Gamma_{\alpha_{k}^{k+1}} (\beta_{k}^{k+1} )   \\
      =\sum_{\sigma \in S_{n}} sgn(\sigma)
                \Gamma_{\alpha_{1} \sigma(1)} \cdots
                \Gamma_{\alpha_{k} \sigma(k)}
                (\beta_{1}, \ldots , \beta_{k},0)   \\
                                                \\
       =\sum_{\sigma \in S_{k}} sgn(\sigma)
        \underbrace{\sum_{\nu_{1}+ \ldots + \nu_{k+1}=\alpha_{1}}
    \cdots \sum_{\mu_{1}+ \ldots + \mu_{k+1}=\alpha_{k}}}_{k \ \text{sums}}
        (\beta_{1}+ \nu_{1}+ \cdots + \mu_{1}, \ldots ,
                0+ \nu_{k+1}+\cdots + \mu_{k+1} ) \label{ksums}
\end{multline}

\begin{eqnarray}  \label{appearsin}
         =\sum_{\sigma \in S_{k}} sgn(\sigma)
        \sum \cdots \sum (\text{signatures whose last entry is zero}) \\
                                                \nonumber \\
        +\sum_{\sigma \in S_{k}} sgn(\sigma)
        \sum \cdots \sum (\text{signatures whose last entry is non-zero})
        \nonumber
\end{eqnarray}
 But the sum (\ref{appearsin}) is just the spectrum of
 $(\alpha)^{k}_{k} \otimes (\beta)^{k}_{k}$ appearing in the spectrum of
  $ (\alpha)^{k+1}_{k} \otimes (\beta)^{k+1}_{k}$.
Finally, the requirement that $0 \leq \nu_{i+1} \leq m_{i}-m_{i+1}$
guarantees
that the application of a simple multiplier to a signature
$(m_{1},\ldots,m_{l},0,\ldots,0)$ extends the length of the signature by
at most one, since $0 \leq \nu_{l+1} \leq (m_{l+1}-m_{l+2})=0$. Thus,
since there are only $k$ sums in (\ref{ksums}),
application of a compound multiplier corresponding to a signature of
length $k$
decomposes the tensor product into a spectrum of signatures of length at
most $l+k$, proving that the spectrum stabilizes.

\end{proof}

\section{A reciprocity theorem}

        According to \cite{ttone} we have the following theorem regarding
{\em dual representations} of Bargmann-Segal-Fock spaces.
\begin{thm}
        Let \[ G_{1} \subset G_{2} \subset  \cdots  \subset G_{k}
                \subset G_{k+1} \subset \cdots \]
be a chain of compact classical groups.  Let $G_{\infty}$ denote the
inductive limit of the $G_{k}$'s.  Let $R_{G_{\infty}}$ and $R'_{G'}$
be given dual representations on $\F_{n \times \infty}$.  Let $H_{\infty}$ be
the inductive limit of a chain of compact subgroups
        \[  H_{1} \subset H_{2} \subset  \cdots  \subset H_{k}
                \subset H_{k+1} \subset \cdots \]
with $H_{k} \subset G_{k}$, and let $R_{H_{\infty}}$ be the representation of
$H_{\infty}$ on $\F_{n \times \infty}$ obtained by restricting
$R_{G_{\infty}}$ to $H_{\infty}$.  If there exists a group $H' \supset G'$
and a representation $R'_{H'}$ on $\F_{n \times \infty}$ such that $R'_{H'}$
is the dual to $R_{H_{\infty}}$ and $R'_{G'}$ is the restriction of
$R'_{H'}$ to the subgroup $G'$ of $H'$ then we have the following
multiplicity free decompositions of $\F_{n \times \infty}$ into isotypic
components
 \[ \F_{n \times \infty} =
        \sum_{(\lambda)} \oplus \I_{n \times \infty}^{(\lambda)}
       = \sum_{(\mu)} \oplus \I_{n \times \infty}^{(\mu)} \]
where ($\lambda$) is a common irreducible signature of the pair
$(G',G_{\infty})$ and $(\mu)$ is a common signature of the pair
$(H',H_{\infty})$.

   If $\lambda_{G_{\infty}}$ (resp. $\lambda'_{G'}$) denotes an irreducible
unitary representation of class $(\lambda)$ and
        $\mu_{H_{\infty}}$ (resp. $\mu'_{H'}$)
denotes an irreducible unitary representation of class

$(\mu),$ then the multiplicity
  \[ \dim \left[ \Hom_{H_{\infty}}
        (\mu_{H_{\infty}} :
         \lambda_{G_{\infty}\left|_{H_{\infty}} ) \right.} \right] \]
of the irreducible representation $\mu_{H_{\infty}}$ in the restriction to
$H_{\infty}$ of the representation $\lambda_{G_{\infty}}$ is equal to the
multiplicity
  \[ \dim \left[ \Hom_{G'}
        (\lambda'_{G'} :
         \mu'_{H'\left|_{G'}  \right.}) \right] \]
of the irreducible representation $\lambda'_{G'}$ in the restriction of the
representation $\mu'_{H'}$.
\end{thm}

   Note that $G'$ and $H'$ are finite dimensional Lie groups and that we
have a similar theorem for the pairs $(G', G_{k})$ and $(H', H_{k})$ where
$G'$ and $H'$ remain fixed for all $k$.  It follows that
 \[ \dim \left[ \Hom_{H_{\infty}}
        (\mu_{H_{\infty}} :
         \lambda_{G_{\infty}\left|_{H_{\infty}}  \right.}) \right] \]
remains constant and the spectral decomposition of
$ \lambda_{G_{k}\left|_{H_{k}}  \right.}$ stabilizes for $k$
large.  To apply this theorem to our problem we first let
  \[ G_{\infty} =
        \underbrace{U_{\infty} \times \cdots
        \times  U_{\infty}}_{r \ copies}\]
acting as exterior tensor product representations on
\[ V^{(m_{1})^{\infty}} \otimes \cdots \otimes V^{(m_{r})^{\infty}}
        \subset \F_{n \times \infty}  ,\]
and $G'=U(p_{1}) \times \cdots U(p_{r})$ acting on $\F_{n \times \infty}$.

Then $H_{\infty} = U_{\infty}$ is the interior tensor product
representation on
        \[ V^{(m_{1})^{\infty}} \otimes \cdots \otimes V^{(m_{r})^{\infty}}, \]
and $H' = U(n)$, where recall that $n = p_{1}+p_{2}+ \cdots + p_{r}$.
This gives the multiplicity of the representation of $U_{\infty}$ with
signature $(m)^{\infty}$ in
$(m_{1})^{\infty} \otimes \cdots \otimes (m_{r})^{\infty}$
in terms of the multiplicity of the representation of the corresponding
representations of $U(p_{1}) \times \cdots \times U(p_{r})$ in the
corresponding representation on $U(n)$.  Next we let
        \[ G_{\infty} =
        \underbrace{U_{\infty} \times \cdots
        \times  U_{\infty}}_{r+1 \ copies}\]
acting as exterior tensor product representations on
\[ V^{(m_{1})^{\infty}} \otimes \cdots \otimes V^{(m_{r})^{\infty}}
   \otimes V^{(m \spcheck )^{\infty}} \subset \F_{n \times \infty}  \]
and $G' = U(p_{1}) \times \cdots \times U(p_{r}) \times U(q)$ acting on
$\F_{n \times \infty}$. Then $H_{\infty}=U_{\infty}$ and $H'=U(p,q)$
where $p+q = n, \ p_{1}+ \cdots + p_{r} = p$.  This gives the
multiplicity of the 
representation of $U_{\infty}$ with signature $(0,\ldots,0,\ldots)^{\infty}$
in the tensor product
$(m_{1})^{\infty} \otimes \cdots \otimes (m_{r})^{\infty}
        \otimes (m \spcheck)^{\infty} $
in terms of the multiplicity of the representation with signature
$(m_{1})^{n} \otimes \cdots \otimes (m_{r})^{n}
        \otimes (m \spcheck)^{n} $
of $U(p_{1}) \times \cdots \times U(p_{r}) \times U(q)$ in the holomorphic
discrete series of $U(p,q)$ with signature (lowest highest weight)
$(\underbrace{0,\ldots,0}_{p},\underbrace{0,\ldots,0}_{q}).$  Note that
these two applications of this theorem can be used together to give another
proof of Theorem \ref{sixone} of Section \ref{decomp}.

\section{Realization of the contragredient representation}
\label{contra}

        A representation $\rho$ of any group $G$ on a vector space $V$
induces in
a natural way a representation $\rho^{*}$ (said to be
{\em contragredient} to $\rho$)
on its dual space $V^{*}$
by
        \[ \rho^{*}(g) \phi (v) = \phi (\rho(g^{-1}) v) \ \ \
                \phi \in V^{*}, \ g \in G. \]
In this section, by making a formal change of variable, we are able to
realize $R_{k}^{*}$ (the representation
contragredient to $R_{k}$) as the representation $R_{k} \spcheck$
on a subspace of polynomial functions of the Fock space $\F_{k}^{\spcheck}$.

        Let $\langle \ \  | \ \ \rangle $ be the inner product on the space
$\F_{k}$ given by (\ref{innerprodb}) or the equivalent inner product
(\ref{innerprod}). Then for any $f \in \F_{k}$ and for each
$k =1,2,3,\ldots$ the mapping
        \[ \Phi : \F_{k} \longrightarrow \F_{k}^{*} \]
given by
\begin{equation*} %\label{isomorphism}
        [\Phi f ](h) = \langle h | f \rangle \ \ \ \ h \in \F_{k}
\end{equation*}
is a conjugate linear isomorphism (or {\em anti-isomorphism}) from
$\F_{k}$ onto its dual space $\F_{k}^{*}$, and it is routine to check
(see \cite{kt}) that $\Phi$ intertwines the representations $R_{k}$ and
$R_{k}^{*}$.  It follows that if $(R^{(m)}_{k}, V^{(m)}_{k})$ is an
irreducible representation of $G_{k}$, and if $V^{(m^{*})}_{k} =
\Phi (V^{(m)}) $, then
$(R^{(m^{*})}_{k}, V^{(m^{*})}_{k}) $ is also an irreducible
representation of $G_{k}$.
It is shown in appendix A of \cite{kt}
that the highest weight vector of $(R^{(m)}_{k}, V^{(m)}_{k})$ with highest
weight $(m_{1},m_{2},\ldots , m_{k})$ is mapped to the lowest weight vector
of $(R^{(m^{*})}_{k}, V^{(m^{*})}_{k})$ with weight
$(-m_{1},-m_{2},\ldots , -m_{k})$, and the lowest weight vector of
$(R^{(m)}_{k}, V^{(m)}_{k})$ with weight $(m_{k}, \ldots , m_{1})$
is mapped onto the highest weight vector of
$(R^{(m^{*})}_{k}, V^{(m^{*})}_{k})$ with weight $(-m_{k}, \ldots , -m_{1})$.
We will realize
$(R^{(m^{*})}_{k}, V^{(m^{*})}_{k})$ on a Fock space $\F_{k}^{\spcheck}$ as
the
representation
$(R^{(m) \spcheck}_{k}, V^{(m \spcheck)})$ constructed as follows.

        Define $(\C^{n \times k}) \  \spcheck$ as the vector space of
complex $n \times k $ matrices
with the reverse ordering
        \[ w =   \left( \begin{array}{cccc}
                                w_{n,k} & \cdots& \ & w_{n,1}\\
                                \vdots & \  & \ & \vdots \\
                                 w_{1,k}  & \cdots& \  & w_{1,1}
                                \end{array} \right)
                                        \in (\C^{n \times k} ) \ \spcheck   \]
Let $s =s_{k}$ be the $k \times k$ matrix with ones along the off
diagonal and
zeros elsewhere

\begin{equation}  \label{s}
        s= \left( \begin{array}{ccccc}
                                0 & 0 & \cdots & 0 & 1  \\
                                 0 & \  & \  & \cdot  & 0  \\
                                 \vdots &  \  & \cdot & \  & \vdots  \\
                                0 & \cdot  & \  & \   &  0 \\
                                1 & 0  & \cdots  & 0  &   0
                                \end{array} \right)
                                        \in \C^{k \times k }
\end{equation}
For future reference one easily checks that if we set
\[ Z =   \left( \begin{array}{cccc}
                                w_{11} & \cdots& \ & w_{1k}\\
                                \vdots & \  & \ & \vdots \\
                                 w_{n1}  & \cdots& \  & w_{nk}
                                \end{array} \right) \]
that $W= s_{n} Z s_{k}$ and that $s=s^{-1}=  s^{T}$.  We then let
$\F_{k} \spcheck $ be the Hilbert space of holomorphic
square-integrable functions on $(\C^{n \times k} ) \ \spcheck$
and define the action $\Rcheck_{k}$ of $G_{k}$ on
 $\F_{k} \spcheck$
 as
     \[ [\Rcheck_{k}(g)f](w) = f(w (sgs) \spcheck)=f(w (s \gcheck s) ). \]
The embedding of $ (\C^{n \times k}) \ \spcheck$ into
$ (\C^{n \times (k + 1) } )\ \spcheck$ given by

\[   W \mapsto \left( \begin{array}{cccc}
                                0 & \  & \ & \  \\
                                \vdots & \ & W & \  \\
                                0 & \ & \ & \
                                \end{array} \right)
                                \in \C^{n \times (k + 1)\ \spcheck}  \]
then induces an embedding $i_{k+1}^{k \ \spcheck}:
        \F_{k} \spcheck \longrightarrow \F_{k+1} \spcheck $,
 we have commutativity of the diagram
\begin{equation*}
\begin{CD}
        G_{k} \times \F_{k}\spcheck     @>R_{k}^{\spcheck}>> \F_{k}\spcheck \\
        @Vj_{k+1}^{k}\times i_{k+1}^{k  \ \spcheck}VV
                        @VVi_{k+1}^{k \ \spcheck}V \\
        G_{k+1} \times \F_{k+1} \spcheck   @>>R_{k+1}^{\spcheck}>
                        \F_{k+1} \spcheck
\end{CD}
\end{equation*}
as in (\ref{cda}), and the inductive limit  representation
$\Rcheck = \varinjlim \, \Rcheck_{k}$ of $G$ on
$\F_{\infty} \spcheck  = \varinjlim \F_{k}\spcheck $
is well defined as in (\ref{welldef}).

        We remark here that the space $\F_{\infty} \spcheck$
  defined above  and the space $\F_{\infty}$
defined in Section \ref{prelim}, are certainly equal {\em as sets}, but are
somewhat different as algebraic
objects, being induced by different embeddings.  In what follows,
the arguments presented in
developing properties of the various inductive limit representations are
readily
modified to any situation.

Let $\fmax$ be the highest weight vector of $(R^{(m)}_{k}, V^{(m)}_{k})$ with
highest weight $(m_{1}, m_{2}, \ldots , m_{k})$.  Then by definition
   \[R(d) \fmax(Z) = d_{11}^{m_{1}} \cdots d_{kk}^{m_{k}} \fmax (Z)
        = \pi^{(m_{1}, m_{2}, \ldots , m_{k})}(d) \fmax (Z) .\]
Define $\fmin^{\spcheck} (W) := \fmax(s_{n}W s_{k}) = \fmax(Z)$, so that
 $ \fmax(Z) = \fmin^{\spcheck} (s_{n}Zs_{k})$.  Then
\begin{alignat*}{2}
 \Rcheck_{k}(d)\fmin^{\spcheck}(W) &= \fmin^{\spcheck}(W s d^{\spcheck} s)\\
 \ &    = \fmax(s_{n}(W  \ s d^{\spcheck} s ) s),
        \ \ \text{since $s_{n}Ws = Z$ and $d^{\spcheck}=d^{-1}$} \\
 \ &    = \fmax(Z d^{-1} ) \\
\ &     =  \pi^{(m_{1}, m_{2}, \ldots , m_{k})}( d^{-1}) \fmax (Z) \\
\ &     =  \pi^{(-m_{1}, -m_{2}, \ldots , -m_{k})}(d) \fmax (Z) \\
\ &     = \pi^{(-m_{1}, -m_{2}, \ldots , -m_{k})}(d) \fmin^{\spcheck}
                (s_{n}Zs_{k})\\
\ &     = \pi^{(-m_{1}, -m_{2}, \ldots , -m_{k})}(d) \fmin^{\spcheck} (W)
\end{alignat*}

        And for $\zeta \in \Zminus$
\begin{alignat*}{2}
\Rcheck_{k}(\zeta)\fmin^{\spcheck}(W)
        &= \fmin^{\spcheck}(W s \zeta^{\spcheck} s)\\
 \ &    = \fmax(s_{n} (W \ s \zeta^{\spcheck} s)s) \\
 \ &    = \fmax(Z \zeta^{\spcheck}) \\
\ &     = \fmax(Z) \\
\ &     = \fmin^{\spcheck}(W) \\
\end{alignat*}
        since if $\zeta \in \Zminus$,  then $\zeta^{\spcheck} \in \Zplus $.
Thus $\fmin^{\spcheck}$ is a lowest weight vector for the representation
$\Rcheck_{k}$, so if we let
$ V_{k}^{(m^{\spcheck})}$ be the  $G_{k}$-submodule generated
by the action $\Rcheck_{k}$ on $ \fmin^{\spcheck}$, then
 $(\Rcheck_{k}, V_{k}^{(m^{\spcheck})})$ is an irreducible representation
of $G_{k}$ characterized by its lowest weight
$(-m_{1}, -m_{2}, \ldots , -m_{k})$,  and it follows that this
representation is
equivalent to the contragredient representation on the dual space
$(R^{(m^{*})}_{k}, V^{(m^{*})}_{k}) $, with
the same lowest weight $(-m_{1}, -m_{2}, \ldots , -m_{k})$.  Furthermore, if
$b \in B_{n}$, and if we set $\tilde{b} = s_{n}bs_{n}$ then, using an
argument similar to the one above, we see that
  \[ \fmin^{\spcheck} (\tilde{b} W ) =  \pi^{(m)} (b) \fmin^{\spcheck} (W) \]
and thus the space $(\Rcheck_{k}, V_{k}^{(m^{\spcheck})})$ can be
characterized as the subspace of polynomial functions that transform
covariantly with respect to the Borel subgroup, as in (\ref{pi}).
We remark here that we refer to $\fmin^{\spcheck}$ as a {\em lowest weight
vector} because it is invariant under right translation by the subgroup
$\Zminus$ which corresponds to the notion of lowest weight using the usual
lexicographic ordering.  It is strictly a matter of choice whether or not to
refer to it as a highest weight vector with respect to the reverse
lexicographic ordering.

     With this realization, if $(m)=(m_{1},m_{2},\ldots, m_{k})$ is the
signature of an
irreducible representation of $G_{k}$, then its contragredient
representation has signature $(m \spcheck)=(-m_{1},-m_{2},\ldots , -m_{k})$,
and it
is routine to check that
\begin{equation} \label{inlimc}
        (-m_{1}, \ldots , -m_{k}) \prec
                (-m_{1}, \ldots , -m_{k}, 0) \prec
                        (-m_{1}, \ldots , -m_{k}, 0, 0) \prec
                                \cdots
\end{equation}
i.e. the appropriate diagram (see (\ref{cdb})) commutes, and so the inductive
limit of the irreducible representations (\ref{inlimc}) is an irreducible
representation of $G_{\infty}$ with signature
 \[ (m \spcheck)^{\infty} =:(-m_{1}, -m_{2}, \ldots , -m_{l}, \zeroinfty) \]
generated by the action $\Rcheck$ on the vector $\fmin^{\spcheck}$.
We will adopt the convention of refering to this as the representation
contragredient to the irreducible representation with signature
$(m)^{\infty}=(m_{1},m_{2},\ldots , m_{k}, \zeroinfty )$, although it is the
{\bf inductive limit} of contragredient representations. We summarize with

 \begin{thm}
If the irreducible representation of
$G_{\infty}$
with signature
 \[(m_{1},m_{2},\ldots , m_{l}, \zeroinfty )\]
 is the
inductive limit of the representations
  \[(m_{1},m_{2},\ldots , m_{l})\preceq (m_{1},m_{2},\ldots , m_{l},0)
        \preceq (m_{1},m_{2},\ldots , m_{l},0,0)\preceq \cdots  \]
then the inductive limit of contragredient representations
 \[(m_{1},m_{2},\ldots , m_{l})^{*}
        \preceq (m_{1},m_{2},\ldots , m_{l},0)^{*}
           \preceq (m_{1},m_{2},\ldots , m_{l},0,0)^{*} \preceq \cdots  \]
is an also irreducible representation of $G_{\infty}$ with signature
        \[ (-m_{1}, -m_{2}, \ldots , -m_{l}, \zeroinfty) .\]
\end{thm}

        We illustrate this idea with the following example.  For each
$k=1,2,3,\dots$ consider $\F_{k}=\F(\C^{1 \times k})$. If
    \[ f^{(m)}(Z) = z_{1}^{m},  \ \ \ \ \ Z=(z_{1}, z_{2}, \ldots, z_{k})
                \in \C^{1 \times k}  \]
it is easy to check that
  \[ f^{(m)}(Zd)= d_{11}^{m} z_{1}^{m} = \pi^{(m,0,\ldots,0)}(d)f^{(m)}(Z)\]
and
        \[ f^{(m)}(Z \zeta)= f^{(m)}(Z), \ \ \zeta \in \Zplus \]
so that $f^{(m)}(Z)$ is a highest weight vector of the representation,
with highest weight $(\underbrace{m,0,\ldots 0}_{k})$.  Right
translation of $ f^{(m)}$ by $G_{k}$ generates the finite dimensional
vector space $P^{(m)} (\C^{1 \times k})$,  of homogeneous polynomials of
degree $m$, so that
$V^{(m,0,\ldots , 0)}= P^{(m)} (\C^{1 \times k}) $ is an irreducible
representation of
$G_{k}$ with signature $(\underbrace{m,0,\ldots 0}_{k})$.  Now
$ P^{(m)} (\C^{1 \times k})$ embeds isometrically into
$ P^{(m)} (\C^{1 \times (k+1)})$, which is also generated as a
$G_{k+1}$- module by right translation of the highest weight vector
$z_{1}^{m}$, which now has highest weight
$(\underbrace{m,0,\ldots 0}_{k+1})$.
Taking the inductive limit of the irreducible representations
   \[ (m) \prec (m,0) \prec \ldots \prec (\underbrace{m,0,\ldots 0}_{k})
        \prec (\underbrace{m,0,\ldots 0}_{k+1}) \prec \ldots \]
we obtain $V^{(m,\zeroinfty)}$, the irreducible representation of $G_{\infty}$
with signature \\
$(m, \zeroinfty) = (m,0,0,\ldots)$,
which is  realized in $\F_\infty$ as the subspace of homogeneous polynomials
of degree $m$, generated by the action $R$ of $G_{\infty}$ on the highest
weight vector $ f^{(m)}(Z)=z_{1}^{m}$;
        \[ V^{(m,\zeroinfty)}= P^{(m)} (z_{1}, z_{2},\ldots) , \ \
                Z \in \C^{1 \times \infty} \]
Now if $w = (w_{k}, \ldots , w_{2}, w_{1})  \in \C^{1 \times k \ \spcheck}$
set $f^{(-m)}(w)= w_{1}^{m}$.\\
If $d= \mathrm{diagonal}\  (d_{11},\ldots,d_{kk}) \ \in \DD_{k}$, then
  \[ \begin{aligned}
        \left[ \Rcheck_{k} (d)f^{(-m)} \right] (w)
           & = \,f^{(-m)}(w (sd \spcheck s) )\\
         \ & =(d_{11}^{-1} w_{11})^{m}  = d_{11}^{-m} f^{(-m)}(w) =
           \pi^{(-m,0,\ldots,0)}(d)f^{(-m)}(Z)
        \end{aligned}  \]
and if $\zeta \in \Zminus_{k}$ then
  \[ \begin{aligned}
    \left[ \Rcheck_{k} (\zeta) f^{(-m)} \right] (w)
        &  =\,f^{(-m)}(w (s \zeta \spcheck s)) \\
         \  & =(\zeta_{11} w_{11})^{m}  = f^{(-m)}(Z) \ \ \
                        \mathrm{since}\  \zeta_{11} =1
     \end{aligned}  \]
Thus $f^{(-m)}$ is a lowest weight vector for the representation
$\Rcheck_{k}$
with lowest weight $\overbrace{(-m,\ldots,0)}^{k}$, and since this holds for
all $k$, we denote the signature of the inductive limit of the
representations
 \[ (-m) \prec (-m,0) \prec \ldots \prec (\underbrace{-m,0,\ldots 0}_{k})
        \prec (\underbrace{-m,0,\ldots 0}_{k+1}) \prec \ldots \]
by $(-m,0,\ldots) = (-m,\zeroinfty)$, and the irreducible
representation
$V^{(-m, \zeroinfty)}$ is realized as the space of homogeneous
polynomials of degree $m$ on $ \C^{1 \times \infty  \spcheck }$, generated
by $f^{(-m)}$.

        \[ V^{(-m,\zeroinfty)}= P^{(m)} (\ldots, w_{2},w_{1}) , \ \
                w \in \C^{1 \times \infty  \spcheck } \]

\section{Decomposing tensor products of irreducible representations}
        \label{decomp}

We now use this construction to realize the tensor product of inductive
limits of irreducible representations.  For
\[  Z^{i}
  = \left( \begin{array}{cccc}
                z_{11}^{i} & z_{12}^{i} & \cdots & z_{1k}^{i}   \\
                \vdots & \ & \   & \vdots \\
                z_{p_{i}1}^{i} & z_{p_{i}2}^{i} & \cdots & z_{p_{i}k}^{i}  \\
        \end{array} \right)
        \in \C^{p_{i} \times k}  \]
set

\[ \left( \begin{array}{c}
                                Z  \\
                                W  \\
                                \end{array} \right)
=\left( \begin{array}{c}
                Z^{1} \\
                Z^{2} \\
                \vdots \\
                Z^{r} \\
                 W
        \end{array}  \right)
  = \left( \begin{array}{cccc}
                                z_{11}  & z_{12}  & \cdots & z_{1k}    \\
                                \vdots & \ & \   & \vdots \\
                                z_{p1}  & z_{p2} & \cdots & z_{pk}  \\
                                   w_{qk}  & \cdots  & w_{q2}& w_{q1} \\
                                \vdots & \    & \      & \vdots \\
                                   w_{1k} & \cdots & w_{12} & w_{11}
                                \end{array} \right)
        \in   \C^{p \times k} \oplus \C^{q \times k \spcheck}  \]
where $p_{1}+ \cdots + p_{r}=p$ and $p+q=n$.
For economy of notation, we now let $\F_{k}$ be the set of holomorphic
square integrable functions on
$\C^{p \times k} \oplus \C^{q \times k \spcheck}$ and define a
representation of $G_{k}$ on $\F_{k}$ by
\begin{equation} \label{rrcheck}
    [R_{k} \otimes \Rcheck_{k}(g) f
        ]\left( \genfrac{(}{)}{0pt}{}{Z}{W} \right)=
   f \left( \genfrac{(}{)}{0pt}{}{Z\, g}{W \,(s g \spcheck s) } \right)
\end{equation}

We then obtain the inductive limit representation $R \otimes \Rcheck $
 of the group $G_{\infty}$ on $\F_{\infty} = \overline{\varinjlim \F_{k}} $
as the representation
induced by the embedding of
\[ \C^{p \times k} \oplus \C^{q \times k \spcheck}
 \longrightarrow
        \C^{p \times (k +1)} \oplus \C^{q \times (k +1) \spcheck}\]
given by

\begin{equation} \label{embed}
                         \left( \begin{array}{c}
                                Z  \\
                                W  \\
                                \end{array} \right)
  \mapsto \left( \begin{array}{ccccc}
                                z_{11} & z_{22} & \cdots & z_{1k}& 0  \\
                                \vdots & \ & \  & \ & \vdots \\
                                z_{p1} & z_{p2} & \cdots & z_{pk}  & 0 \\
                                   0 & w_{qk}  & \cdots & w_{q2} & w_{q1} \\
                                \vdots & \    & \ & \      & \vdots \\
                                   0   & w_{1k} & \cdots & w_{12} & w_{11}
                                \end{array} \right)
        \in   \C^{p \times (k+1)} \oplus \C^{q \times (k+1) \spcheck}
\end{equation}
and the embedding of $G_{k} \longrightarrow G_{k+1}$ given by (\ref{G}).

        If
$(m)^{i} = \overbrace{(m_{1}^{i}, m_{2}^{i}, \ldots , m_{p_{i}}^{i}
, 0 , \ldots,  0}^{k} )$ is the signature of an irreducible representation
of $G_{k}$,
and if
$(m) \spcheck = (-m_{1}, -m_{2}, \ldots ,-m_{q},0, \ldots , 0 )$ is
the signature of the representation
contragredient to $(m) = ( m_{1},m_{2}, \dots, m_{q},0 , \ldots, 0 )$
we form the $n$-tuple of positive integers
\begin{equation} \label{mu}
   \mu = (m_{1}^{1}, m_{2}^{1}, \ldots , m_{p_{1}}^{1},
        m_{1}^{2},   \ldots , m_{p_{2}}^{2},
        \ldots m_{1}^{r},   \ldots , m_{p_{r}}^{r},
         m_{1}, m_{2}, \ldots ,m_{q} )
\end{equation}

        If $B_{i}$, $i=1,\ldots ,r$ is the Borel subgroup of lower
triangular matrices of
$GL(p_{i}, \C)$ and if $B_{q}$ is the Borel subgroup of  $GL(q, \C)$\ ,
for
$ b \in B_{q}$ we first set $ \tilde{b}=sbs$, where $s=s_{q}$ as in (\ref{s}),
and then set $\tilde{B_{q}} = \{ \tilde{b} \ | \ b \in B_{q} \} $.
The group
$B_{1} \times B_{2} \times \cdots \times B_{r} \times \tilde{B_{q}}$ can
then be
identified with the group of all lower triangular block matrices
$\beta$ of the form

\begin{equation} \label{beta}
\beta =  \left( \begin{array}{ccccc}
                \boxed{b_{1}} & \ & \  & \ & \\
                \     & \boxed{b_{2}} & \ &  0  & \ \\
                \      &  \           & \ddots & \   &  \  \\
                \       &  0   & \     & \boxed{b_{r}} & \  \\
        \        & \           & \        &   \    & \boxed{\tilde{b}}
          \end{array}  \right)  \ \ \ \ b_{i} \in B_{i}, \  b \in B_{q}
\end{equation}
where $p_{1}+ \cdots + p_{r}=p$ and $p+q=n$.
It is a consequence of the Borel-Weil Theorem (see for example \cite{kt})
 that for $k \geq n$ the
tensor product of
irreducible $G_{k}$ modules
\begin{equation} \label{tprod}
 V^{(m^{1})^{k}} \otimes \ldots  \otimes V^{(m^{r})^{k}} \otimes
                V^{(m \spcheck)^{k}}
\end{equation}
with the $G_{k}$-action given by
(\ref{rrcheck}),
can be realized as the subspace of polynomial functions $f \in \F_{k}$
which, using the terminology of this paper, satisfy the covariant condition
  \begin{equation} \label{covariantb}
     f \left( \beta \, \genfrac{(}{)}{0pt}{}{Z}{W} \right)=
        \pi^{(\mu)}(\beta) f \left( \genfrac{(}{)}{0pt}{}{Z}{W} \right)
  \end{equation}
for $\mu$ as in (\ref{mu}) and where $\pi^{(\mu)}(\beta) =
        (b_{1})_{11}^{m_{1}^{1}} \cdots (b)_{qq}^{m_{q}}$,
  as in (\ref{pi}).  Since this covariant condition holds for all $k \geq n$,
we realize the tensor product of irreducible $G_{\infty}$-modules
\begin{equation*}
   V^{(m^{1})^{\infty}}\otimes \ldots \otimes V^{(m^{r})^{\infty}}
        \otimes V^{(m \spcheck)^{\infty}}
\end{equation*}
 as the
inductive limit of irreducible $G_{k}$-modules (\ref{tprod}) induced by
the embeddings (\ref{embed}) and (\ref{G}),
whose elements transform according to the covariant condition
(\ref{covariantb}).

For each $k$, let $I^{k}$ denote the identity representation of $G_{k}$
appearing in the tensor product
\begin{equation*}
 V^{(m^{1})^{k}}\otimes \cdots \otimes V^{(m^{r})^{k}}
        \otimes V^{(m \spcheck)^{k}}.
\end{equation*}
then $I^{k}$ has signature $(\underbrace{0,\ldots , 0}_{k} )$ and by
definition there exists a non-zero element
  \[ f_{k} \in V^{(m^{1})^{k}}\otimes \cdots \otimes V^{(m^{r})^{k}}
        \otimes V^{(m \spcheck)^{k}} \]
such that $[R_{k} \otimes R_{k}^{\spcheck} ](g) f_{k} = f_{k} $ for all
$g \in G_{k}$.  This means that $f_{k}$ is {\em invariant} under the action
$R_{k} \otimes R_{k}^{\spcheck}$ of $G_{k}$.  Now it is well known from the
theory of invariants (see for example \cite{we}) that the algebra of
polynomial invariants under this $G_{k}$ action is generated by the $pq$
algebraically independent polynomial functions
\begin{equation} \label{pijk}
  P_{\alpha \beta}^{k}(Z,W)=(Z\,sW^{T})_{\alpha \beta}
     =\sum_{t=1}^{k} Z_{\alpha,t} W_{\beta,t}
\ \ \  1 \leq \alpha \leq p , \ 1 \leq \beta \leq q.
\end{equation}
By our realization of the $V^{(m^{i})^{k}}$ and $V^{(m \spcheck)^{k}}$ as
$G_{k}$-modules we obviously have the isometric embedding
\[ V^{(m^{1})^{k}}\otimes \cdots \otimes V^{(m^{r})^{k}}
        \otimes V^{(m \spcheck)^{k}}
           \subset
                V^{(m^{1})^{k+1}}\otimes \cdots \otimes V^{(m^{r})^{k+1}}
        \otimes V^{(m \spcheck)^{k+1}} \]
of $G_{k}$-modules into $G_{k+1}$-modules. It is routine to check that
the appropriate diagrams commute, and as in (\ref{cda}) and (\ref{cdb})
we obtain the
representation $R \otimes R^{\spcheck}$ of $G_{\infty}$
on $ V^{(m^{1})^{\infty}}\otimes \cdots \otimes V^{(m^{r})^{\infty}}
 \otimes V^{(m \spcheck)^{\infty}}$ as an inductive limit of representations
of $G_{k}$, $k=1,2,3,\ldots$.  But the case of the identity representation is
entirely different.  For each $k$ let $\I^{k}$ denote the one-dimensional
subspace of $ V^{(m^{1})^{k}}\otimes \cdots \otimes V^{(m^{r})^{k}}
        \otimes V^{(m \spcheck)^{k}}$ spanned by the invariant vector $f_{k}$
mentioned above.  Then we obviously can not define the inductive limit of
$I^{k}$.  However we can define the {\em inverse} or {\em projective limit}
of the family $\{ G_{k}, I^{k}, \I^{k} \}$ as follows:
For each pair of indices $j, k$ with $j \leq k$ define a continuous
homomorphism $\phi_{j}^{k}: \I^{k} \longrightarrow \I^{j}$ such that
  \begin{align}
    a) \ & \phi_{j}^{j} \  \text{is the identity map for all $j$,} \notag \\
    b) \ & \text{if $i \leq j \leq k$, then
                  $\phi_{i}^{k} =\phi_{j}^{k} \circ \phi_{i}^{j}$}. \notag
        \end{align}
Here we can take $\phi_{j}^{k}$ as the {\em truncation homomorphism}, i.e.
$\phi_{j}^{k}$ is defined on the generators $P_{\alpha \beta}^{k}$ by
\begin{equation} \label{truncate}
         \phi_{j}^{k} (P_{\alpha \beta}^{k}) = P_{\alpha \beta}^{j}
  \ \ \ \text{ for $j \leq k$ }
\end{equation}
The {\em inverse limit} of the system  $\{ \I^{k},\phi_{j}^{k} \}$ is then
formally defined by
  \[ \I^{\infty} := \varprojlim \I^{k} =
        \left\{ (f_{k}) \in \prod_{k} \I^{k} \ | \ f_{i}= \phi_{i}^{j}(f_{j})
        \ \ \ \text{whenever $ i \leq j$}  \right\} \]
Concretely we can define the functions
 \begin{equation} \label{pabinf}
        P_{\alpha \beta}:= P_{\alpha \beta}^{\infty}
                =\lim_{k \longrightarrow \infty} P_{\alpha \beta}^{k}
                =\sum_{t=1}^{\infty} Z_{\alpha,t} W_{\beta,t} \ \
\ \ \  1 \leq \alpha \leq p , \ 1 \leq \beta \leq q
\end{equation}
and make the following observations for each $\alpha, \beta$:

   1)  $P_{\alpha, \beta}$ is well defined on
  \[  \C^{p \times \infty} \oplus \C^{q \times \infty \spcheck}
   =\bigcup_{k=1}^{\infty}\left( \C^{p \times k} \oplus
                \C^{q \times k \spcheck}\right) \]

  2)$P_{\alpha, \beta}$ is \underline{not} an element of $\F_{\infty}$, but
instead lies in
   $\varprojlim \ \F_{k}$, the {\em projective limit} or {\em inverse limit} of the
 of Bargmann-Segal-Fock spaces
$\F_{k}$ (for details on the projective limit representations of
$G_{\infty}$ see \cite{rmh}).

      It follows that any $f \in \I^{\infty} $ has the form
\begin{equation} \label{prod}
        f= \sum C_{IJK} \prod (P_{\alpha \beta})^{\gamma}
\end{equation}
where the functions $P_{\alpha \beta}$ are as defined in (\ref{pabinf})
for $1 \leq \alpha \leq p, \  \ 1 \leq \beta \leq q $,
the $\gamma$ are non-negative integers,
the sums and products in (\ref{prod}) are finite, and the $C_{IJK}$
are constants
with multi-indices $I,J$ and $K$. Let
$\pi_{k}:\I^{\infty} \longrightarrow \I^{k}$ denote the projection of
$\I^{\infty}$ onto $ \I^{k}$.  Let $I^{\infty}$ denote the representation of
$G_{\infty}$ on  $\I^{\infty}$ given by the following equation
 \begin{equation} \label{thirtyone}
   I^{\infty} (g) f =
     \sum C_{IJK} \prod \lim_{k \longrightarrow \infty}
       \left[ \left( R \otimes R^{\spcheck}(g)
        P_{\alpha \beta}^{k} \right)^{\gamma} \right]  \ \
                \text{for $g \in G_{\infty}$ and $ f \in \I^{\infty}$ }
\end{equation}
Since $g \in G_{\infty}$ means that $g \in G_{j}$ for some $j$, and for
$k \geq j$
  \[  \left[ R \otimes R^{\spcheck}(g) \right]
        P_{\alpha \beta}^{k}=P_{\alpha \beta}^{k} \]
equation (\ref{thirtyone}) implies that $P_{\alpha \beta}$ are $G_{\infty}$-
invariant, and hence $I^{\infty} (g) f = f$ for all $ f \in \I^{\infty}$.
It follows that $\pi_{k}\left(I^{\infty} (g) f \right) = \pi_{k} (f)$ for all
$ g \in G_{\infty}$ and $ f \in \I^{\infty}$.

        Recall that if $\PP_{k} = \PP(\C^{n \times k})$ denotes the subspace
of all polynomial functions of $\C^{n \times k}$ then $\PP_{k}$ is dense in
$\F_{k}$.  Let
        \[ \PP_{\infty} = \bigcup_{k=1}^{\infty} \ \PP_{k} \]
denote the inductive limit of $\PP_{k}$, then clearly $\PP_{\infty}$ is
dense in $\F_{\infty}$. Let  $\PP_{\infty}^{*}$ (resp.  $\F_{\infty}^{*}$)
denote the {\em dual} or {\em adjoint} space of $\PP_{\infty}$
(resp. $\F_{\infty}$).  Then since $\PP_{\infty}$ is dense in $\F_{\infty},$
 $\F_{\infty}^{*}$ is dense in $\PP_{\infty}^{*}$.  By the Riesz
representation theorem for Hilbert spaces, every element
$f^{*} \in \F_{\infty}^{*}$ is of the form $\langle \cdot \ | f \rangle$
for some $f \in \F_{\infty}$, and the map $f^{*} \mapsto f$ is an anti-linear
 (or conjugate-linear) isomorphism.  Thus we can identify $\F_{\infty}^{*}$
with $\F_{\infty}$ and obtain the rigged Hilbert space as the triple
$\PP_{\infty} \subset \F_{\infty} \subset \PP_{\infty}^{*}$ (see \cite{gs}
for the definition of rigged Hilbert spaces). Typically and element
$P_{\alpha \beta}$ defined by equation (\ref{pabinf}) belongs to
$\PP_{\infty}^{*}$, and if $ f \in \F_{\infty}$ then $f \in \F_{k}$ for some
$k$, so we can define the inner product
\begin{equation}   \label{ippabf}
  \langle P_{\alpha \beta} , f \rangle
        = \langle \pi_{k} (P_{\alpha \beta}), f \rangle
           = \langle P_{\alpha \beta}^{k} , f \rangle
\end{equation}
in fact, in the calculation of
   \[\left.  P_{\alpha \beta}(D) \overline{f(\bar{Z})} \right|_{Z=0} \]
the terms in $P_{\alpha \beta}$ whose column indices are larger than $k$ drop
 off.

\begin{thm} \label{sixone}
        Let $V^{(m^{1})^{\infty}},\ldots ,V^{(m^{r})^{\infty}}$ and
        $ V^{(m)^{\infty}}$ be irreducible representations of $G_{\infty}$.
Using the convention of Section
\ref{contra}, let $ V^{(m \spcheck)^{\infty}}$ be the representation
contragredient to $ V^{(m)^{\infty}}$.
Let $I^{\infty}$ be the identity representation defined by Equation
\ref{thirtyone}.
Then the multiplicity of $ V^{(m)^{\infty}}$ in the tensor product
        \[ V^{(m^{1})^{\infty}}\otimes \ldots \otimes V^{(m^{r})^{\infty}}\]
is equal to the multiplicity of $I^{\infty} $ in the tensor product
        \[ V^{(m^{1})^{\infty}}\otimes \ldots \otimes V^{(m^{p})^{\infty}}
                \otimes V^{(m \spcheck)^{\infty}} .\]
\end{thm}
\begin{proof}
From \cite{kt} we know that for sufficiently large $k$ the multiplicity of
$V^{(m)^{k}}$ in
  \[ V^{(m^{1})^{k}}\otimes \ldots \otimes V^{(m^{r})^{k}}\]
is equal to the multiplicity of the identity representation $I^{k}$ in the
augmented tensor product
 \[ V^{(m^{1})^{k}}\otimes \ldots \otimes V^{(m^{r})^{k}}
                \otimes V^{(m \spcheck)^{k}} .\]
For each $k$ let $h_{k}$ denote the homomorphism sending the irreducible
representation of $G_{k}$ with signature $(0, \ldots, 0)$ into the
$G_{k}$-module
        \[ V^{(m^{1})^{k}}\otimes \ldots \otimes V^{(m^{r})^{k}}
                \otimes V^{(m \spcheck)^{k}} .\]
Then
        \[ \phi_{j}^{k} \circ  h_{k} = h_{j} \ \ \text{ for $j \leq k$} \]
where the homomorphisms $\phi_{j}^{k}$ are defined as in (\ref{truncate}).
Let $(0, \ldots , 0)^{\infty}$ denote the signature of the representation of
$G_{\infty}$ as the \underline{inverse limit} of irreducible representations
of $G_{k}$ with signature $\underbrace{(0, \ldots , 0 )}_{k} $.  Then we can
define a homomorphism
   \[ h: V^{(0, \ldots , 0)^{\infty}} \longrightarrow
         V^{(m^{1})^{\infty}}\otimes \ldots \otimes V^{(m^{r})^{\infty}}
                \otimes V^{(m \spcheck)^{\infty}} \]
by
\begin{equation} \label{h}
        h(v) = \varprojlim \ h_{k} \left( \pi_{k} (v) \right)
\end{equation}
where in Equation (\ref{h}), $\pi_{k}$ denotes the projection of
 $  V^{(0, \ldots , 0)^{\infty}} $ onto
$ V^{(0, \ldots , 0)^{k}}  $.  Note that $ V^{(0, \ldots , 0)^{k}}$
or $V^{(0, \ldots , 0)^{\infty}}$ are just the trivial $G_{k}$ or
$G_{\infty}$ modules $\C$, and that the $G_{\infty}$-module
   \[  V^{(m^{1})^{\infty}}\otimes \ldots \otimes V^{(m^{r})^{\infty}}
                \otimes V^{(m \spcheck)^{\infty}} \]
is considered as a $G_{\infty}$-submodule of the $G_{\infty}$-module
$\PP_{\infty}^{*}$.  As remarked in Section \ref{stability}, the dimension
of
  \[ \Hom_{G_{k}} \left( V^{(0,\ldots , 0)^{k}} ,
        V^{(m^{1})^{k}}\otimes \ldots \otimes V^{(m^{r})^{k}}
                \otimes V^{(m \spcheck)^{k}}  \right) , \]
the space of all homomorphisms intertwining $V^{(0,\ldots , 0)^{k}}$ and
$V^{(m^{1})^{k}}\otimes \ldots \otimes V^{(m^{r})^{k}}
                \otimes V^{(m \spcheck)^{k}} $
stabilizes as $k$ gets large.  But this dimension is just the multiplicity
of $I^{k}$ in $ V^{(m^{1})^{k}}\otimes \ldots \otimes V^{(m^{r})^{k}}
                \otimes V^{(m \spcheck)^{k}} $
which, in turn is equal to the multiplicity of $ V^{(m)^{k}}$ in
$ V^{(m^{1})^{k}},\ldots ,V^{(m^{r})^{k}}$.
It follows that at the (inductive) limit we have
\begin{multline*}
  \dim \left[ \Hom_{G_{\infty}} \left( V^{(0,\ldots , 0)^{\infty}} ,
        V^{(m^{1})^{\infty}}\otimes \ldots \otimes V^{(m^{r})^{\infty}}
                \otimes V^{(m \spcheck)^{\infty}}  \right) \right] \\
= \dim  \left[ \Hom_{G_{\infty}} \left( V^{(m)^{\infty}} ,
        V^{(m^{1})^{\infty}}\otimes \ldots \otimes V^{(m^{p})^{\infty}}
                 \right) \right]
\end{multline*}
or equivalently
 the multiplicity of $ V^{(m)^{\infty}}$ in the tensor product
        \[ V^{(m^{1})^{\infty}}\otimes \ldots \otimes V^{(m^{r})^{\infty}}\]
is equal to the multiplicity of $I^{\infty} $ in the tensor product
        \[ V^{(m^{1})^{\infty}}\otimes \ldots \otimes V^{(m^{r})^{\infty}}
                \otimes V^{(m \spcheck)^{\infty}} .\]
\end{proof}

 Let $\{f^{m^{i}}_{\xi_{i}} \}_{\xi_{i}} $ be a basis of state vectors for
$V^{(m^{i})^{\infty}}$ ,
$ i = 1 \ldots r$, let
$\{f^{m}_{\xi} \}_{\xi} $ be a basis of state vectors for $V^{(m)^{\infty}}$
and let
$\{f^{m^{*}}_{\xi^{*}} \}_{\xi^{*}} $ be
a basis of state vectors for $V^{(m \spcheck)^{\infty}}$.
Then
\[f^{m^{1}}_{\xi_{1}} \otimes f^{m^{2}}_{\xi_{2}} \otimes
        \ldots \otimes f^{m^{r}}_{\xi_{r}} \]
is a natural basis for the tensor product of irreducible representations
\begin{equation*}
   V^{(m^{1})^{\infty}}\otimes \ldots \otimes V^{(m^{r})^{\infty}}
\end{equation*}
and
\[f^{m^{1}}_{\xi_{1}} \otimes f^{m^{2}}_{\xi_{2}} \otimes
        \ldots \otimes f^{m^{r}}_{\xi_{r}} \otimes f^{m^{*}}_{\xi^{*}} \]
is a natural basis for the tensor product of irreducible representations

\begin{equation*}
   V^{(m^{1})^{\infty}}\otimes \ldots \otimes V^{(m^{r})^{\infty}}
        \otimes V^{(m \spcheck)^{\infty}}
\end{equation*}

        Let $\{ \I_{\eta} \}_{\eta}$ be a basis for the
$G_{\infty}$-invariant subspace which is `{\em contained}' in
\begin{equation*}
   V^{(m^{1})^{\infty}}\otimes \ldots \otimes V^{(m^{r})^{\infty}}
        \otimes V^{(m \spcheck)^{\infty}}
\end{equation*}
 in the sense described above.
If we set
\[ \I_{\eta}  \left( \left( \begin{array}{c}
                                Z  \\
                                W  \\
                                \end{array} \right) \right)
                = \I_{\eta}(Z,W) \]
 and consider $\I_{\eta}(Z,W)$
 as a function of $W$, and also note that any function
$f \in V^{(m \spcheck)^{\infty}}$ is a function of $W$ alone, then we can
form the
inner product, as defined in (\ref{ippabf})
\begin{equation}  \label{ipw}
   \langle \I_{\eta} \ | \ f \rangle_{W}
         = \I_{\eta}(Z,D) \overline{f (\bar{W})}|_{W=0}
\end{equation}
and thereby obtain a function of $Z$.

Considering the
 remarks above, we adapt the statement and proof of
 Theorem 2.3 of \cite{kt}, to our situation as follows
\begin{thm}  \label{sixtwo}
        Let
        \[ \tilde{f}_{\xi}^{m,\eta} (Z)
   = \langle \I_{\eta}(Z,W) \  | \ {f}_{\xi^{*}}^{m^{*}}(W) \rangle_{W}
         = \I_{\eta}(Z,D) \overline{f_{\xi^{*}}^{m^{*}}(\bar{W})}|_{W=0}  \]

Then $ \{\tilde{f}_{\xi}^{m,\eta} \}_{\xi}$ is an isomorphic image of
$\{f^{m}_{\xi} \}_{\xi}$ in
$V^{(m^{1})^{\infty}}\otimes \ldots \otimes V^{(m^{r})^{\infty}}$ indexed
by the multiplicity label $\eta$ and we have the following relation of
Clebsch-Gordan coefficients
\begin{equation} \label{cgcs}
 \langle  \tilde{f}_{\xi}^{m,\eta}  |
  f^{m^{1}}_{\xi_{1}} f^{m^{2}}_{\xi_{2}} \ldots f^{m^{r}}_{\xi_{r}} \rangle
= \langle  \I_{\eta}  | f^{m^{1}}_{\xi_{1}} f^{m^{2}}_{\xi_{2}} \ldots
f^{m^{r}}_{\xi_{r}}
f^{m^{*}}_{\xi^{*}} \rangle
\end{equation}
\end{thm}
\begin{proof}
To first show that $ \tilde{f}_{\xi}^{m,\eta} (Z)$ in fact lies in
$V^{(m^{1})^{\infty}}\otimes \ldots \otimes V^{(m^{r})^{\infty}}$
it is sufficient to show (by the Borel-Weil theorem)  that if
$b=(b_{1},\ldots , b_{r}) \in B_{1} \times \cdots \times B_{r}$ then, as in
Equation (\ref{covariantb})
  \[  \tilde{f}_{\xi}^{m,\eta} (bZ)
         = \pi^{\mu (m^{1})}(b_{1}) \cdots \pi^{\mu (m^{r})}(b_{r})
                 \tilde{f}_{\xi}^{m,\eta} (Z)    \]
But since $\I_{\eta}$ `{\em lies} ' in
\begin{equation*}
   V^{(m^{1})^{\infty}}\otimes \ldots \otimes V^{(m^{r})^{\infty}}
        \otimes V^{(m \spcheck)^{\infty}}
\end{equation*}
it transforms covariantly with respect to the Borel subgroup 
defined in Equation (\ref{beta}) so we have
\[ \begin{aligned}
  \tilde{f}_{\xi}^{m,\eta} (bZ)
 = & \,\langle \I_{\eta}\ (bZ,W) \
        | \ f_{\xi^{*}}^{m^{*}}(W) \rangle_{W} \\
 = & \,  \langle \I_{\eta} \ (bZ,Id \ W) \  |
         \ f_{\xi^{*}}^{m^{*}}(W) \rangle_{W}
        \ \ \text{ where $Id$ is the $q \times q$ identity matrix} \\
 = & \  \langle \ \pi^{\mu}(\beta) \  \I_{\eta}(Z, \,W) \  |
  \ f_{\xi^{*}}^{m^{*}}(W) \rangle_{W} \ \
        \text{where $ \beta =b \times Id$} \\
 =& \  \pi^{\mu}(\beta) \  \langle \ \I_{\eta}(Z, \,W) \  |
  \ f_{\xi^{*}}^{m^{*}}(W) \rangle_{W} \ \ \text{the inner product
        is linear in the first argument} \\
 =& \  \pi^{\mu (m^{1})}(b_{1}) \cdots \pi^{\mu (m^{r})}(b_{r})
                 \tilde{f}_{\xi}^{m,\eta} (Z)
                  \ \  \text{by Equation (\ref{covariantb}), as desired.}
\end{aligned}\]

We next show that the $ \{\tilde{f}_{\xi}^{m,\eta} \}_{\xi}$ transform under
the representation $R^{(m)}$ in the same manner as the
$ \{ f_{\xi}^{m} \}_{\xi}$. Since $\I_{\eta}(Z,W)$ is invariant with respect
to the action $R \otimes R^{\spcheck}$ of $G_{\infty}$ we have
  $\I_{\eta}(Zg,W) = \I_{\eta}(Z,W s g^{ -1 \spcheck} s) $ which can 
succinctly be
written as $R(g)\I_{\eta}(Z,W)= R^{\spcheck}(g^{-1}) \I_{\eta}(Z,W)$.
We also have that
  \[ R^{(m)}(g) f_{\xi}^{m} = \sum_{\xi'} \D^{m}_{\xi \xi'}(g) f_{\xi'}^{m} \]
where the $ \D^{m}_{\xi \xi'}$ are the $D$-functions for the representation
$R^{(m)}$.  Now for any $g \in G_{\infty}$ we can assume that $g \in U(k)$
for some $k$, so that $g^{\spcheck} = \bar{g}$.  Hence
 $ \D^{m}_{\xi \xi'}(g^{\spcheck})= \overline{ \D^{m}_{\xi \xi'}(g)}$, and
it follows from the definitions of the symbols involved that
        \[  R^{ (m) \spcheck }(g)  f_{\xi^{*}}^{m^{*}} = \sum_{\xi^{*'}}
           \overline{ \D^{m}_{\xi \xi'}(g)} f_{\xi^{*'}}^{m^{*}}   \]
Thus we seek to show that
 \[ R^{(m)}(g) \tilde{f}_{\xi}^{m,\eta}
    = \sum_{\xi'} \D^{m}_{\xi \xi'}(g) \tilde{f}_{\xi'}^{m,\eta} \]
By the preceding remarks and the definition of $\tilde{f}_{\xi}^{m,\eta}$
we then have
\[ \begin{aligned}
  R^{(m)}(g) \tilde{f}_{\xi}^{m,\eta}(Z)
        = & \,\langle \I_{\eta}\ (Zg,W) \
        | \ f_{\xi^{*}}^{m^{*}}(W) \rangle_{W} \\
        = & \,\langle R^{(m)}(g) \, \I_{\eta}\ (Z,W) \
        | \ f_{\xi^{*}}^{m^{*}}(W) \rangle_{W} \\
        = & \,\langle  R^{(m) \spcheck} (g^{-1}) \, \I_{\eta}\ (Z,W) \
        |  f_{\xi^{*}}^{m^{*}}(W) \rangle_{W} \\
        = & \,\langle \, \I_{\eta}\ (Z,W) \
        | \ R^{(m) \spcheck} (g){f}_{\xi^{*}}^{m^{*}}(W) \rangle_{W}
               \ \  \ \ \text{since the representation is unitary} \\
        = & \,\langle \I_{\eta}\ (Z,W) \
        | \ \sum_{\xi^{*'}} \overline{\D^{m}_{\xi \xi'}(g)}
                f_{\xi^{ *'}}^{m^{*}}(W) \rangle_{W} \\
        = & \,\sum_{\xi'} \D^{m}_{\xi \xi'}(g)
                \, \langle \I_{\eta}\ (Z,W) \
                | \ {f}_{\xi^{*'}}^{m^{*}}(W) \rangle_{W}
                \ \ \ \ \ \text{by conjugate linearity} \\
        = & \, \sum_{\xi'} \D^{m}_{\xi \xi'}(g)
                 \tilde{f}_{\xi'}^{m,\eta}
\end{aligned} \]

Finally we have
\[ \begin{aligned}
 \langle  \I_{\eta}  |
 f^{m^{1}}_{\xi_{1}} f^{m^{2}}_{\xi_{2}} \ldots f^{m^{r}}_{\xi_{r}}
    f^{m^{*}}_{\xi^{*}} \rangle
        = & \, \langle  \I_{\eta}  |
 f^{m^{*}}_{\xi^{*}}f^{m^{1}}_{\xi_{1}} f^{m^{2}}_{\xi_{2}}
    \ldots f^{m^{r}}_{\xi_{r}} \rangle \\
        = & \, \I_{\eta} (D,D)
         \overline{f^{m^{*}}_{\xi^{*}}(\bar{W})} \,
          \overline{f^{m^{1}}_{\xi_{1}} (\bar{Z})}
          \ldots
           \overline{f^{m^{r}}_{\xi_{r}}(\bar{Z})}|_{(Z,W)=(0,0)}\\
       = & \, \left[ \I_{\eta} (D,D)
         \overline{f^{m^{*}}_{\xi^{*}}(\bar{W})} \right] \,
          \overline{f^{m^{1}}_{\xi_{1}} (\bar{Z})}
          \ldots
           \overline{f^{m^{r}}_{\xi_{r}}(\bar{Z})}|_{(Z,W)=(0,0)}\\
       = & \, \tilde{f}_{\xi}^{m,\eta}(D)
          \overline{f^{m^{1}}_{\xi_{1}} (\bar{Z})}
           \ldots \overline{f^{m^{r}}_{\xi_{r}}(\bar{Z})}|_{Z=0} \\
        = & \, \langle  \tilde{f}_{\xi}^{m,\eta}  |
            f^{m^{1}}_{\xi_{1}} f^{m^{2}}_{\xi_{2}} \ldots
              f^{m^{r}}_{\xi_{r}} \rangle
\end{aligned}  \]
which is Equation (\ref{cgcs}).
\end{proof}

\section{Example}

       We illustrate the techniques described in this paper with the
example
 \[ (7,1,\zeroinfty) \subset
(1,\zeroinfty) \otimes (2,\zeroinfty) \otimes (2,\zeroinfty)
        \otimes (3,\zeroinfty) \]
considered in (\ref{gfive}) of  Section \ref{stability}.  By the results of
Theorem \ref{sixtwo}  and Equation (\ref{covariantb}) we seek
algebraically independent polynomials of the form
\begin{equation} \label{pfourtwo}  P   \left( \genfrac{(}{)}{0pt}{}{Z}{W} \right)
        = \sum C_{IJK} \prod (P_{\alpha \beta})^{\gamma}
                \ \ \ \alpha = 1,2,3,4 \ \ \ \beta = 1,2
\end{equation}
that satisfy the covariant condition
\begin{equation} \label{Pbeta}
     P \left( \beta \, \genfrac{(}{)}{0pt}{}{Z}{W} \right)=
        \pi^{(\mu)}(\beta) f \left( \genfrac{(}{)}{0pt}{}{Z}{W} \right)
\end{equation}
where $\mu = (1,2,2,3,7,1)$ and
\begin{equation*}
\beta =  \left( \begin{array}{cccccc}
            b_{1} & \ & \  & \ & \ & \\
            \     & b_{2} & \ &  \  & 0 & \ \\
            \     &  \           & b_{3} & \   &  \ & \  \\
            \     &  \   & \     &  b_{4} & \  & \ \\
         \  & \       0    & \        &   \    &  b_{5}& b* \\
         \  & \           & \        &   \    &   & b_{6}
          \end{array}  \right)  \ \ \ \ b_{i}, \, b*  \in C.
\end{equation*}

If $\DD$ is the diagonal subgroup and $\Zplus$ is the upper triangular
unipotent subgroup, then  $\beta \in \DD \Zplus$ so we can first
reduce the problem by solving (\ref{Pbeta}) for the diagonal subgroup
$\DD$ which consists of elements of the form
\begin{equation*}
 d =  \left( \begin{array}{cccccc}
            b_{1} & \ & \  & \ & \ & \\
            \     & b_{2} & \ &  \  & 0 & \ \\
            \     &  \           & b_{3} & \   &  \ & \  \\
            \     &  \   & \     &  b_{4} & \  & \ \\
         \  & \       0    & \        &   \    &  b_{5}& \ \\
         \  & \           & \        &   \    &   & b_{6}
          \end{array}  \right)  \ \ \ \ b_{i}  \in \C.
\end{equation*}
Hence we seek polynomials of the form
\begin{equation*} \label{poly}
        P=P_{11}^{\ell_{11}}
           P_{12}^{\ell_{12}}P_{21}^{\ell_{21}}
                \cdots P_{41}^{\ell_{41}}
                     P_{42}^{\ell_{42}}
\end{equation*}
 that satisfy

\begin{equation*}
     P \left( d \genfrac{(}{)}{0pt}{}{Z}{W} \right)=
        \pi^{(\mu)}(d) f \left( \genfrac{(}{)}{0pt}{}{Z}{W} \right)\, , 
		\forall \ d \in \DD.
\end{equation*}
This leads us to the system
\[ \begin{cases}
        \ell_{11}+\ell_{12}=1 \\
        \ell_{21}+\ell_{22}=2 \\
        \ell_{31}+\ell_{32}=2 \\
        \ell_{41}+\ell_{42}=3 \\
        \ell_{11}+\ell_{21}+\ell_{31}+\ell_{41}=7 \\
        \ell_{12}+\ell_{22}+\ell_{32}+\ell_{42}=1
\end{cases} \]
which gives us the following set of polynomials that transform covariantly
with respect to the diagonal subgroup $\DD$;
\[ \begin{aligned}
        P_{1}=& P_{11}P_{21}P_{22}P_{31}^{2}P_{41}^{3} \\
        P_{2}=& P_{11}P_{21}^{2}P_{31}P_{32} P_{41}^{3} \\
        P_{3}=& P_{11}P_{21}^{2}P_{31}^{2}P_{41}^{2}P_{42} \\
        P_{4}=& P_{12}P_{21}^{2} P_{31}^{2}P_{41}^{3}.
\end{aligned} \]

Next, from (\ref{pfourtwo}) and (\ref{Pbeta}) we seek functions of the form
\begin{equation}  \label{thetaP}
           P = C_{1} P_{1} +
                 C_{2} P_{2} +
                     C_{3} P_{3} +
                        C_{4} P_{4}
\end{equation}
that transform covariantly with respect to the upper triangular unipotent
subgroup $\Zplus$ which consists of elements of the form
\begin{equation*}
Z^{+} =  \left( \begin{array}{cccccc}
            1 & \ & \  & \ & \ & \\
            \     & 1 & \ &  \  & 0 & \ \\
            \     &  \           & 1 & \   &  \ & \  \\
            \     &  \   & \     &  1 & \  & \ \\
         \  & \       0    & \        &   \    &  1& b* \\
         \  & \           & \        &   \    &   & 1
          \end{array}  \right)  \ \ \  b*  \in \C.
\end{equation*}
Checking this condition on $P_{1},\ P_{2}, \ P_{3}$ and  $ P_{4}$ we see that
\[ \begin{aligned}
         P_{1} \left( Z^{+} \genfrac{(}{)}{0pt}{}{Z}{W} \right)
          = \ & P_{1} + b*P_{11} P_{21}^{2} P_{31}^{2} P_{41}^{3} \\
        & \\
         P_{2}  \left( Z^{+} \genfrac{(}{)}{0pt}{}{Z}{W} \right)
          = \ & P_{2} + b*P_{11} P_{21}^{2} P_{31}^{2} P_{41}^{3} \\
        & \\
         P_{3} \left( Z^{+} \genfrac{(}{)}{0pt}{}{Z}{W} \right)
          = \ & P_{3} + b*P_{11} P_{21}^{2} P_{31}^{2} P_{41}^{3} \\
        & \\
         P_{4} \left( Z^{+} \genfrac{(}{)}{0pt}{}{Z}{W} \right)
          =  & P_{4} + b*P_{11} P_{21}^{2} P_{31}^{2} P_{41}^{3} 
\end{aligned} \]

In order that 
	\[  P\left( Z^{+} \genfrac{(}{)}{0pt}{}{Z}{W} \right)
	   =  P\left( \genfrac{(}{)}{0pt}{}{Z}{W} \right) \ 
		\ \ \forall Z^{+} \in \Zplus  \]
we must have
$C_{1}+C_{2}+C_{3}+C_{4}=0$. Thus a convenient basis
of $G_{\infty}$-invariants in this tensor product can be chosen as
\[ \begin{aligned}
        \I_{1} =& \  P_{1}-P_{2}
                =& P_{11}P_{21}P_{22}P_{31}^{2}P_{41}^{3}
                        - P_{11}P_{21}^{2}P_{31}P_{32} P_{41}^{3} \\
        \I_{2} =& \  P_{2}-P_{3}
                 =& P_{11}P_{21}^{2}P_{31}P_{32} P_{41}^{3}
                         -P_{11}P_{21}^{2}P_{31}^{2}P_{41}^{2}P_{42} \\
        \I_{3} =& \ P_{3}-P_{4}
                 =& P_{11}P_{21}^{2}P_{31}^{2}P_{41}^{2}P_{42}
                         - P_{12}P_{21}^{2} P_{31}^{2}P_{41}^{3}.
\end{aligned} \]
Note that the space of invariants has dimension three, which is the
multiplicity of $(7,1,\zeroinfty)$ computed earlier.

        Now a natural basis for the $G_{\infty}$-invariant subspace with
signature $(1,\zeroinfty)$  contained in $\F_{\infty}$ as described in
Section \ref{decomp} is given by $\{ Z_{1i} \}_{i=1}^{\infty}$.  Similarly
$\{ Z_{2i}Z_{2j} \}_{i,j=1}^{\infty}, \
         \{ Z_{3i}Z_{3j} \}_{i,j=1}^{\infty}$ and
        $\{ Z_{4i}Z_{4j} Z_{4k}\}_{i,j,k=1}^{\infty}$
are natural basis for the subspaces
$(2,\zeroinfty), \ (2,\zeroinfty)$ and $(3,\zeroinfty)$, respectively,
and an element of $(7,1,\zeroinfty)^{\spcheck}$is its lowest weight
vector
        $\, w_{11}^{6} \det
           \left(
                \begin{smallmatrix}
                   w_{22} & w_{21} \\
                   w_{12} & w_{11}
                \end{smallmatrix} \right) $.
Thus an example of a basis element for the tensor product
 \[ (1,\zeroinfty) \otimes (2,\zeroinfty) \otimes (2,\zeroinfty)
        \otimes (3,\zeroinfty) \otimes  (7,1,\zeroinfty)^{\spcheck} \]
would be
\begin{equation} \label{basis}
  Z_{11} Z_{21}^{2} Z_{31}^{2} Z_{41}^{3}
        (W_{11}^{7}W_{22}-W_{11}^{6}W_{21})
\end{equation}
and to compute a Clebsch-Gordan coefficient we compute the inner product
of (\ref{basis}) with, for example $\I_{1}$.
\begin{multline*} \label{cgc}
  \langle \I_{1} \, | \,Z_{11} Z_{21}^{2} Z_{31}^{2} Z_{41}^{3}
    (W_{11}^{7}W_{22}-W_{11}^{6}W_{21}) \rangle = \\
 \left[ P_{11}P_{21}P_{22}P_{31}^{2}P_{41}^{3}
           - P_{11}P_{21}^{2}P_{31}P_{32} P_{41}^{3} \right] (D) \\
        Z_{11} Z_{21}^{2} Z_{31}^{2} Z_{41}^{3}
         (W_{11}^{7}W_{22}-W_{11}^{6}W_{21})  \bigr|_{(Z,W)=(0,0)}
\end{multline*}
We remark that in the above computation, for example the product
        \[ P_{11}P_{21}P_{22}P_{31}^{2}P_{41}^{3}(D)
        =\left( \sum_{t=1}^{\infty} Z_{1t}W_{1t} \right) \cdots
                \left( \sum_{t=1}^{\infty}  Z_{4t}W_{1t} \right)^{3} (D) \]
need only be evaluated up to $t=2$ since those terms whose column indices
are larger than two evaluate to zero in the above inner product.
  This is routinely accomplished using
a computer algebra system, such as {\em Maple}.

\section{Conclusion}
        We have shown how the multiplicity problem the Clebsch-Gordan
coefficients in the decomposition of r-fold tensor products of irreducible
tame representations of $U(\infty)$ can be restated in terms of
$U(\infty)$-invariants.  Thus all the theorems for $U(k)$ treated in
\cite{kt} can be generalized to $U(\infty)$.  Actually the computational
aspect of the problems are much simpler with this new approach and one can
use computers to obtain invariant polynomials, and by differentiating these
polynomials compute Clebsch-Gordan and Racah coefficients.

\section*{Acknowledgments}
R. M. Howe was partially
supported by the University of Wisconsin-Eau Claire Office of University
Research.

T. Ton-That was partially supported while on
 Developmental Research Assignment at the University of Iowa.

\bibliography{UINF}%\end{thebibliography)

\end{document}